\documentclass{MITcsail}
\usepackage{indentfirst}
\usepackage{subfigure}

\title{Dynamic Viscosity of Methane and Carbon Dioxide Hydrate Systems from Pure Water at High-Pressure Driving Forces}

\author
{Andr\'e Guerra\orcidA{}, Adam McElligott\orcidB{}, Chong-Yang Du, Milan Mari\'c\orcidC{}, Alejandro D. Rey\orcidE{}, Phillip Servio\orcidD{}\footnote{Correspondence E-mail: phillip.servio@mcgill.ca}\\
\vspace{1em} 
\normalfont{\small Department of Chemical Engineering, McGill University, Montr\'eal, QC, Canada}\\
}

\begin{document}

\maketitle
\thispagestyle{firstpagestyle} 

\begin{abstract}
The viscosity of methane and carbon dioxide hydrate systems were measured using a high-pressure rheometer up to 30 MPag. Where hydrate formation was not detected, the effect of temperature on the viscosity was one order of magnitude higher than the pressure effect on viscosity in most of the experimental pressure range (-0.048 mPa$\cdot$s/\textcelsius{} at 1 MPag and 0.009 mPa$\cdot$s/MPag at 2\textcelsius{}). The pressure effect on the viscosity of carbon dioxide systems where no hydrate formation was observed was up to one order of magnitude higher than that of the methane systems, due to carbon dioxide’s higher solubility in water. Novel rheological phases diagrams were developed to further characterize the gas hydrate systems. Several systems with high driving forces for hydrate formation (2.07 MPag to 4.1 MPag) did not form gas hydrates. System limitations to the formation of hydrates were categorized as kinetic, mass diffusion, and/or heat of crystallization effects.
\end{abstract}

\section{Introduction}
Gas hydrates, or clathrate hydrates, form when “guest” gas molecules become trapped in cages formed as water molecules begin to self-organize into a “host” lattice during a liquid-to-solid phase transition \cite{Sloan2008}. These cages are formed by hydrogen bonds between water molecules, and their structure’s potential energy is stabilized by the presence of the guest gas molecules \cite{Sloan2008,Carroll2014}. Sir Humphrey Davy first described natural gas hydrates in 1811 \cite{Davy1811}. However, the research motivation behind hydrate studies remained primarily academic until the early-to-mid 20th century, when the formation of gas hydrates inside pipelines and drill wells began posing operational and economic obstacles in the oil and gas industry. The oil and gas industry continues to be a major driving force in gas hydrate research\cite{Hammerschmidt1934}. As a result, much of the research in the field has focused on inhibitory additives to prevent or slow the formation of gas hydrates in oil and gas systems\cite{Posteraro2015_part1,Posteraro2015_part2,Heidaryan2010,Daraboina2015,Zhukov2017a,Rajput2018,Rajput2021}.

Utilizing the inherent properties of gas hydrates has been suggested for a variety of new technologies. This has added motivation for gas hydrate research beyond oil and gas applications. Gas hydrate properties of interest include high gas-to-water volume ratios (up to 184:1) and the selectivity of guest gas species \cite{Sloan2008,Eslamimanesh2012a}. Some of the suggested hydrate technologies include post-combustion carbon capture from flue gas \cite{Aaron2005,Linga2007,Kang2000a}, storage and transport of natural gas \cite{Gudmundsson1994,Mimachi2015}, water desalination \cite{Park2011}, gas separations \cite{Eslamimanesh2012a,Fan2009a}, two-phase refrigerant applications \cite{Clain2012}, and fruit juice concentration \cite{Loekman2019}. As a result, research into promoter additives that induce hydrate formation, such as graphene nanofluids, has recently become more prominent \cite{McElligott2019,McElligott2021a} . Many of the new hydrate technologies proposed above involve continuous (flowing), or semi-continuous processes. Hydrate formation in these systems may be induced while the system must be kept in flow state. Technologies such as these would benefit from rheological characterization of gas hydrate systems under different thermodynamic conditions. Viscosity data is crucial to guide process design and operational control of such systems.

\begin{figure}[ht]
\centering
\includegraphics[scale = 1]{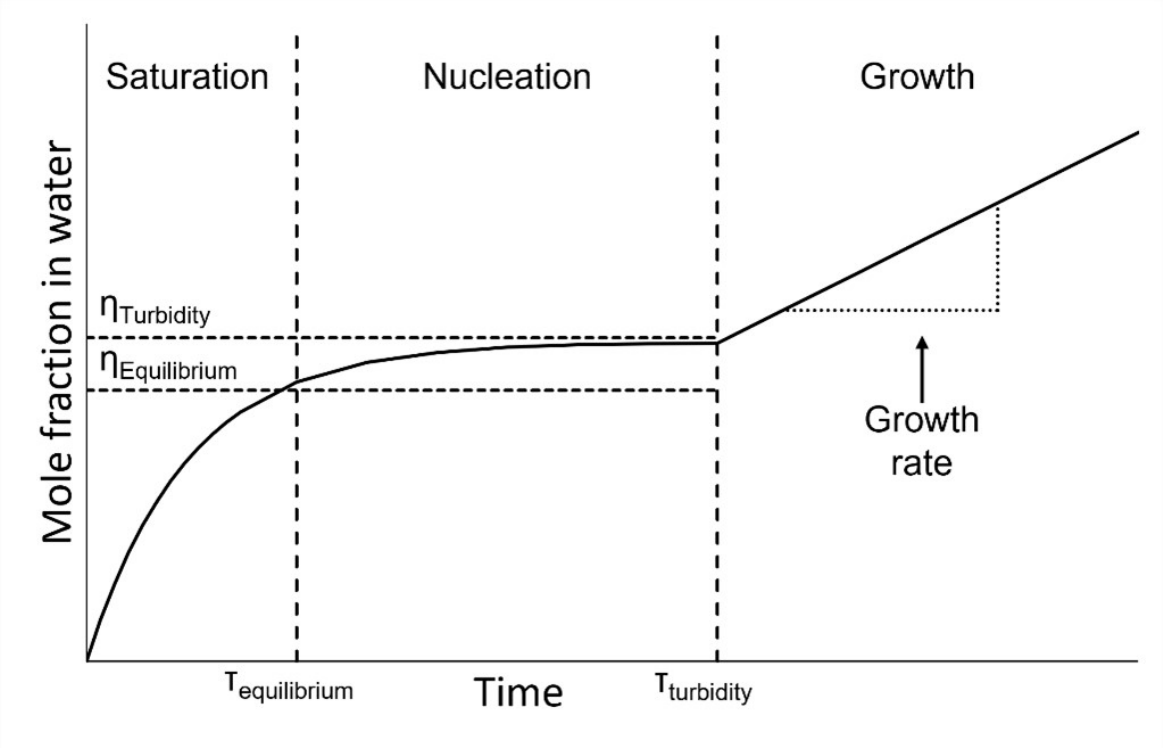}
\caption{Mole fraction of guest gas species in water during the three phases of gas hydrate formation.}
\label{fig:1ghkineticphases}
\end{figure}

Figure~\ref{fig:1ghkineticphases} describes the formation of gas hydrates through three distinct phases: saturation, nucleation, and growth \cite{Sloan2008}. During saturation, the gas species is dissolved in water and its molar fraction in water increases up to an equilibrium concentration ($\eta_{equilibrium}$). If the driving force across the gas–liquid interface is sustained, the gas species continues to dissolve into the water throughout the nucleation phase and eventually the water becomes supersaturated \cite{Sloan2008}. The nucleation process is stochastic, and hydrate nuclei form and dissociate continuously during this phase. At the end of the nucleation phase, the nuclei reach critical mass ($\eta_{turbidity}$) and crystals being to grow starting the growth phase \cite{Sloan2008}. The hydrate growth rate is initially linear and is sustained while physical and thermodynamic conditions promote hydrate formation.

\begin{figure}[ht]
\centering
\includegraphics[scale = 1]{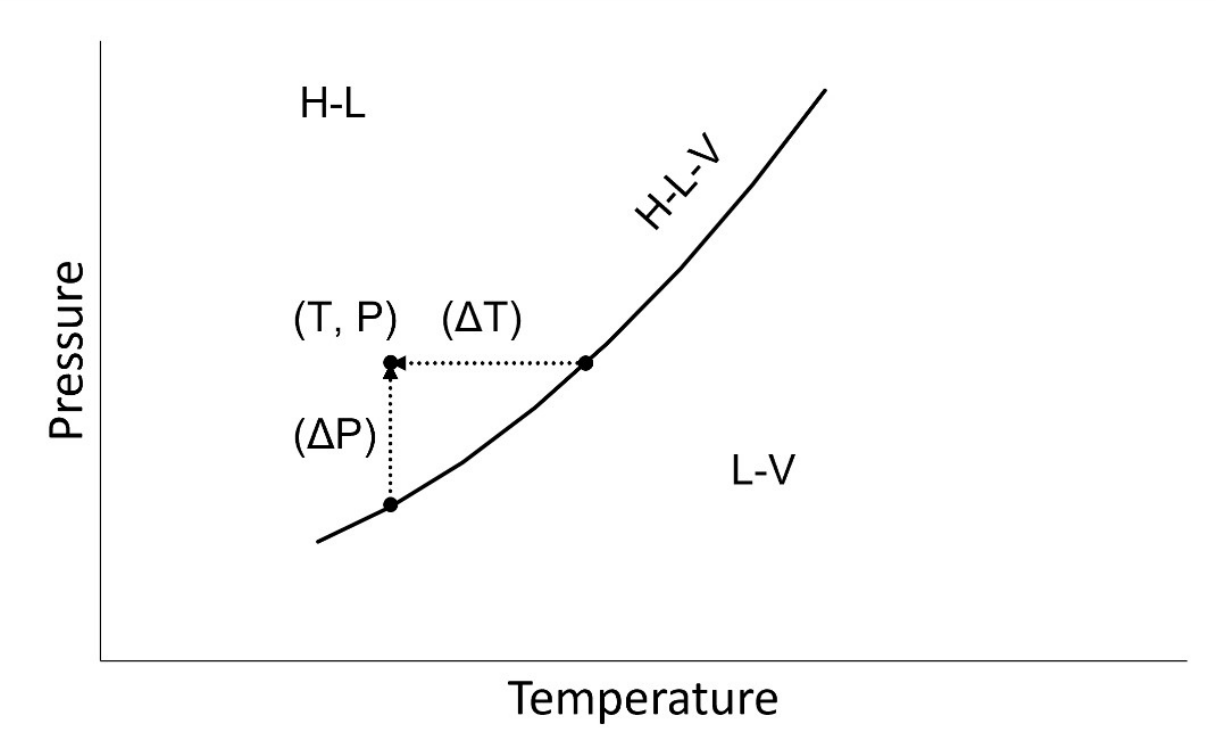}
\caption{Generic representation of the thermodynamic phase diagram of a hydrate forming system.}
\label{fig:2thermophasediag}
\end{figure}

The formation of gas hydrates is governed by the thermodynamic state of the hydrate forming system\cite{Sloan2008}. Hydrate formation occurs at low temperatures and high pressures. Figure~\ref{fig:2thermophasediag} is a generic representation of the phase equilibrium diagram of a hydrate forming system. The diagram contains a three-phase hydrate-liquid-vapour (H-L-V) equilibrium line, which describes thermodynamic conditions where all three phases can coexist. Above and below the H-L-V line, are the hydrate-liquid (H-L) and liquid-vapour (L-V) phase regions, respectively, which indicate conditions where two phases coexist. In applications where hydrate formation is undesirable, e.g., flow systems in the oil and gas industry, it is advantageous to keep the operating point (T, P) below the equilibrium line to prevent hydrate formation. Conversely, in the case of technologies that benefit from hydrate properties, as listed above, the operating point (T, P) should be above the equilibrium line to promote the formation of hydrates. This can be achieved using various types of additives like thermodynamic inhibitors or promoters, respectively. Thermodynamic inhibitors can shift the equilibrium line above the operating point, while thermodynamic promoters can shift it below the operating point. One additional consequence to the use of additives is the change in flow properties of the resulting system. This can affect how the process is designed or operated and thus is a focus of current research.

Rheological characterization of hydrate forming systems has also been motivated by the oil and gas industry and thus focused on water-in-oil hydrate systems\cite{Webb2014,Majid2018,Pandey2017,Pandey2020,Sun2020,Webb2013,Webb2012}. The highest experimental pressures attempted in such research have been between 10-15 MPag due to equipment limitations\cite{Webb2014,Pandey2017,Pandey2020,Webb2012,Webb2013}. Rheological models were developed to predict relative viscosities of hydrate systems such as the Krieger-Dougherty\cite{Krieger1959} and the Mills\cite{Mills1985} models. Both models make use of estimations of the hydrate particle volume fraction, the maximum possible hydrate volume fraction of the system, and assume sphericity of hydrate clusters in suspension. The Camargo-Palermo model was proposed in 2002 and has been widely used to predict the viscosity of oil emulsion hydrate slurries in pipelines and well lines in the oil and gas industry\cite{Camargo2000a,Camargo2002,Majid2018}. This model predicts the gas hydrate particle size and applies the Mills model to determine the relative viscosity of the system. Additionally, the Majid-Wu-Koh model has recently been proposed to predict the viscosity of hydrate systems, in the form of a differential equation, with respect to the volume fraction of hydrate particles, and reports improved performance over the previous models\cite{Majid2017}.

Due to their focus on oil emulsion systems, the models above would be inappropriate for application in other industries. New gas hydrate technologies involve aqueous hydrate forming systems which deviate from the oil systems often explored in literature. The aim of this work is to characterize and elucidate the shear rheology of methane and carbon dioxide hydrate formation in pure water systems. Moreover, the work extends the current literature in gas hydrate rheology to extreme high-pressure conditions (15 to 30 MPag). This work will serve as a baseline for comparisons in future work investigating the rheology of promoted and inhibited hydrate forming systems at extreme high pressures.

\section{Materials and methods}
\begin{figure}[ht]
\centering
\includegraphics[scale = 0.8]{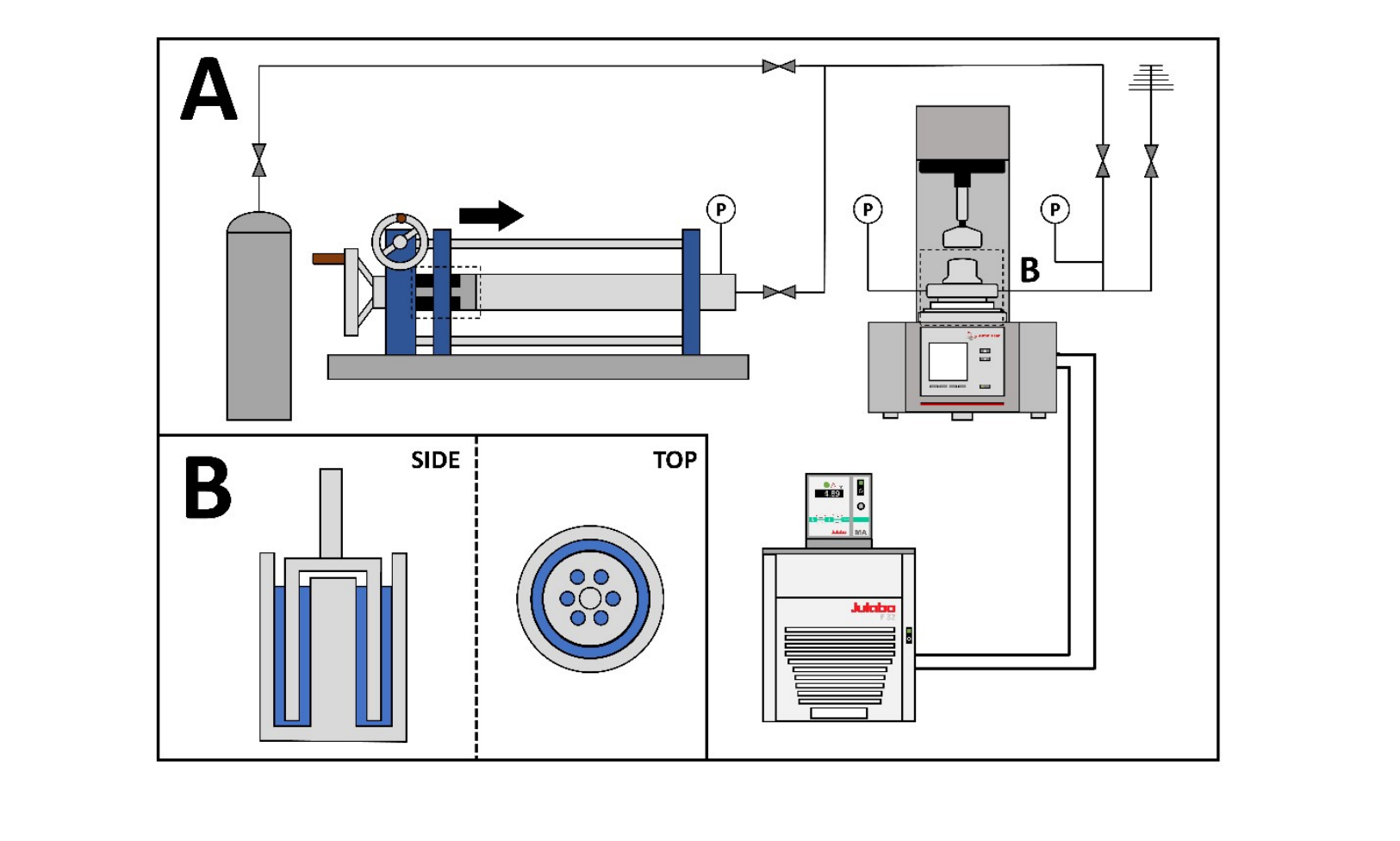}
\caption{Depiction of the experimental setup used in this work; A: gas cylinder, mechanical piston system, Anton Paar MC302 rheometer, and Julabo F32 chiller; B: double-gap (DG) measurement geometry side and top views.}
\label{fig:3rheometer}
\end{figure}

The experiments presented here were conducted in the system depicted in Figure~\ref{fig:3rheometer}. An Anton Paar MCR302 rheometer equipped with a high-pressure (HP) cell was used to collect viscosity measurements. The rheometer’s HP cell has a maximum pressure rating of 40 MPag. A double-gap (DG) measurement geometry and magnetic measurement head were used to conduct these experiments. The measurement head magnetically induced rotational motion on the DG measurement geometry. The DG geometry consists of a double annulus space where the sample was loaded (Figure~\ref{fig:3rheometer}B). The DG geometry was filled with a 7.5 mL sample of reverse osmosis (RO) purified water in all test runs. The RO water was produced with a 0.22 $\mu$m filter, had a conductivity of 10 $\mu$S, and a maximum organic content of 10 ppb. The sample’s temperature was maintained constant using a circulating current Julabo F32 chiller. The refrigerant fluid used was a 50 vol.\% mixture of ethylene glycol in water. Ultrapure (99.99\%) methane and carbon dioxide gas cylinders purchased from MEGS were used in this work. The pressure available from the MEGS gas cylinders was enough to achieve all desired pressure conditions for the carbon dioxide test runs. However, the high pressures explored in the methane test runs (above 10 MPag) required gas compression prior to pressurization of the rheometer. Therefore, a mechanical piston system (Figure~\ref{fig:3rheometer}A) compressed a sample of methane gas in the piston chamber to reach the pressures required for this study.

This work systematically explored the regions below and above H-L-V equilibrium conditions. All methane and carbon dioxide test run conditions included temperatures of 0, 2, 4, 6, 8, and 10\textcelsius{}. At each temperature, the pressures explored in methane test runs were: 0, 1, 2, 3, 4, 5, 10, 15, 20, 25, and 30 MPag, while the carbon dioxide pressure conditions were limited to at most 3 or 4 MPag due to its liquification at higher pressures and lower temperatures. Of the potential test conditions, therefore, carbon dioxide would have entered the liquid phase at 0\textcelsius{} and 4 MPag, and at 2\textcelsius{} and 4 MPag. These conditions were removed from the planned carbon dioxide experimental matrix. Figure 4(a) and (b) list the experimental conditions for methane and carbon dioxide, respectively. Conditions in which hydrate formation is thermodynamically favourable (above the equilibrium line) are indicated with an “x”, while non-favourable conditions (below the equilibrium line) are indicated with an “o”. Additionally, both figures present the driving force for each hydrate forming condition. The equilibrium lines presented in Figure 4(a) and (b) are based on the data presented by Carroll in Natural Gas Hydrates\cite{Carroll2014}. Additionally, conditions with driving forces below 0.5 MPag were considered too low for hydrate formation in the system described above and are presented in red in Figure 4. These conditions were treated as non-hydrate forming for the purposes of this work.

\begin{figure}
     \centering
     \begin{subfigure}
         \centering
        \includegraphics[scale=1]{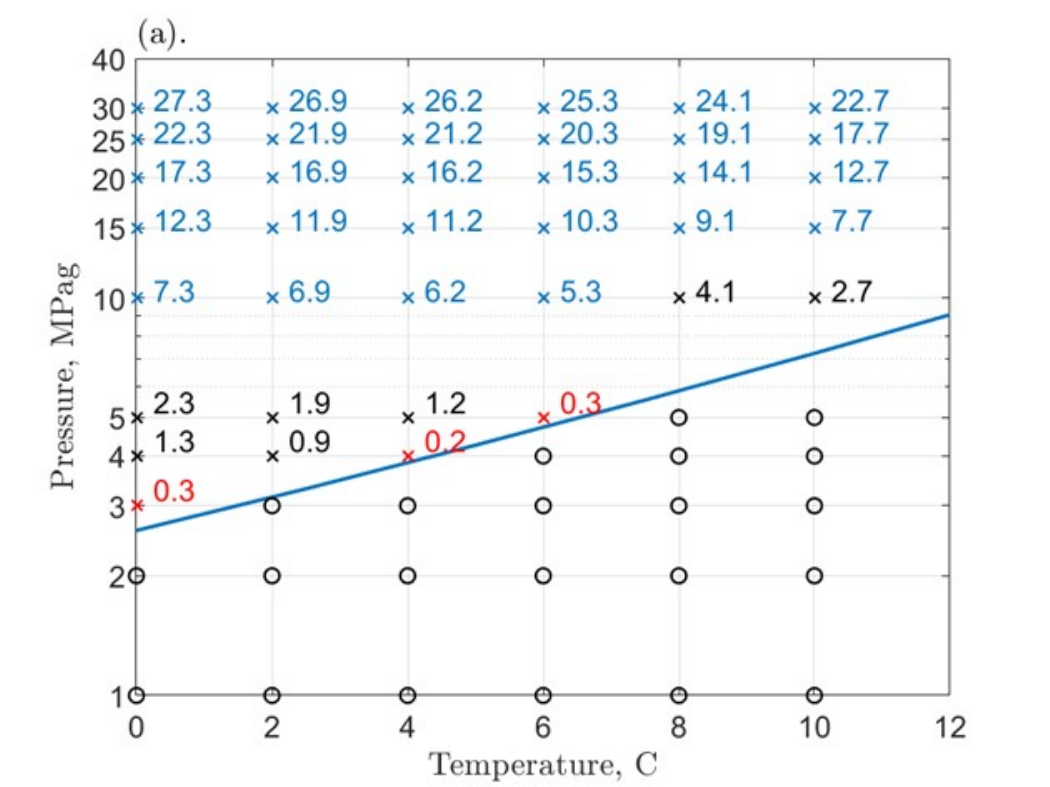}
     \end{subfigure}
     
     \begin{subfigure}
         \centering
         \includegraphics[scale=1]{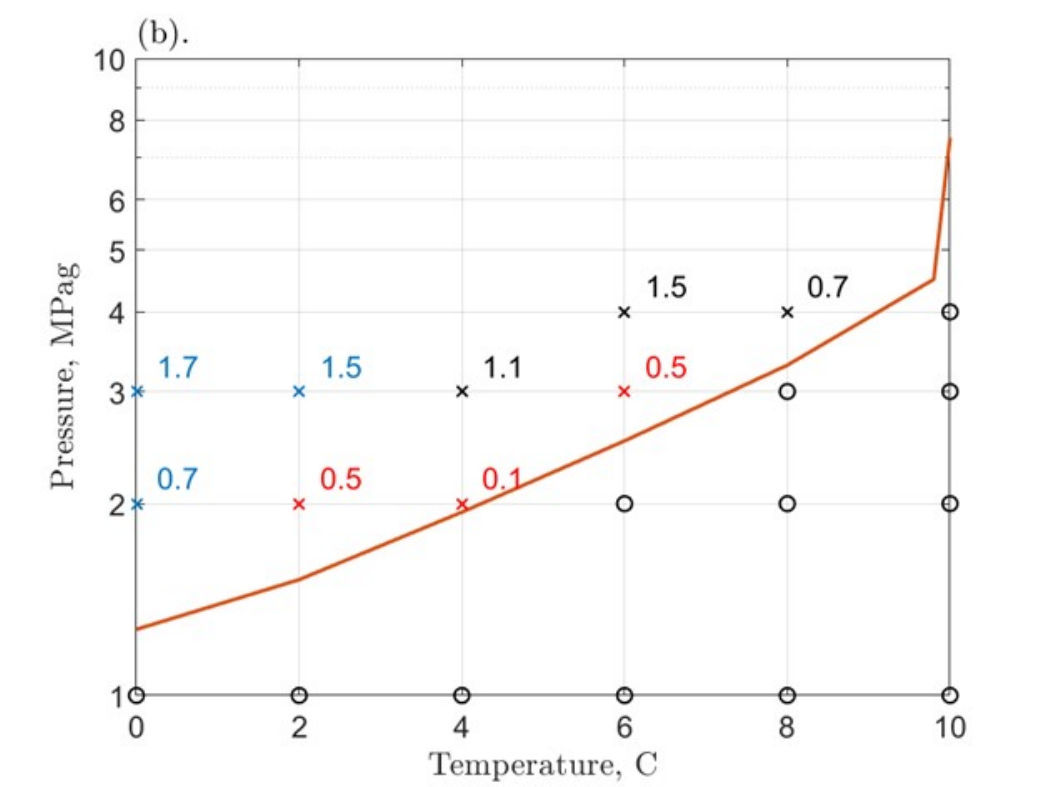}
     \end{subfigure}
     \caption{Thermodynamic phase diagram of methane (a) and carbon dioxide (b); experimental conditions and driving forces are presented in relation to the H-L-V equilibrium line\cite{Carroll2014}; ‘x’: hydrate forming conditions; ‘o’: non-hydrate forming conditions; driving forces are presented in MPag; blue: hydrate formation observed; red: driving forces at or below 0.5 MPag.}
     \label{fig:4expcond}
\end{figure}

Each test run began with calibration of the rheometer’s measurement head inertia, and bearings and motor adjustments were completed as recommended by the rheometer’s manufacturer (Anton Paar). The 7.5 mL sample of RO water was loaded into the well of the HP cell and the DG geometry insert was used to close the cell. The sample head space was purged of air through a repeated dilution method using the gas type associated with the experimental run. For instance, before a methane test run, methane gas was charged into the HP cell at a pressure of approximately 1 MPag. The gas was left stagnant at this pressure for 30 seconds and then let out through the exit valve. This was repeated five times before every test run to ensure all air was purged of the system. Once the sample temperature was stable to +/- 0.1\textcelsius{} of the test run’s temperature setpoint, the rheometer’s measurement system was activated, and the HP cell was then charged instantly to the test pressure with the same gas used for purging. In the case of methane test runs requiring pressures above 10 MPag, the mechanical piston system in Figure~\ref{fig:3rheometer}A was used. The 500-cc piston chamber was pressurized with methane to the maximum available cylinder pressure (approximately 10 MPag) and the position of the piston was then moved to pressurize the gas. The final chamber volume depended on the desired final pressure and the initial charging pressure available from the gas cylinder. 

The rheometer was configured to run at a constant 400 s-1 shear rate in all conditions and for both gases. The shear rate value was selected as it is within the adequate range for the torque measurement magnitude associated with low viscosity liquids such as water, which is the starting state of the system. Additionally, it is the manufacturer recommendation for the use of the DG measurement geometry in the case of low viscosity liquids. It is important to note that the systems examined here utilize the rheometer as both a measurement device and a reactor. As described in the introduction, the formation of gas hydrates is a kinetic process that involves multiple stages (Figure~\ref{fig:1ghkineticphases}). The hydrate formation dynamics introduces a source of variation as the systems progress through the stages throughout each test in which conditions are favourable for hydrate formation. The constant shear rate ensured that temporal viscosity measurements were consistent throughout the gas hydrate formation process. This enabled meaningful discussion of controlled viscosity measurements of the dynamic systems as they progress through the phases of gas hydrate formation. All measures above were intended to reduce as much error as possible from measurements at low viscosities.

All test runs with conditions above the three phase equilibrium lines for methane and carbon dioxide (Figure~\ref{fig:4expcond}) can be classified as hydrate forming conditions due to a positive pressure driving force. All runs considered hydrate forming conditions were allowed a maximum 24-hour period for the maximum viscosity to be reached. Hydrate formation was detected by increased viscosity above the expected water viscosity at the test conditions. Once formation was detected, the test run was continued until the rheometer’s maximum safety limit torque measurement (150 mN$\cdot$m) was reached. At this point the rheometer automatically stopped collecting data, and the test run was ended. Non-hydrate forming condition test runs were allowed to proceed until a continuous ten-minute period of stable viscosity data was collected. This precautionary protocol eliminated the effect of temperature change and gas dissolution on the viscosity measurements at the beginning of test runs. Data collection was performed using the Anton Paar software RheoCompass v.1.25, and data analysis was performed using MATLAB\textregistered{}.

The rheometer’s magnetic measurement head in the HP cell used two ball bearings which were replaced based on wear detected by conducting the rheometer’s template bearing check and motor adjustment tests. The ball bearings were worn due to the length and severeness of the test runs. Both aspects were more prominent during hydrate forming condition runs. On average ball bearings lasted approximately 150 hours of run time with hydrate-forming runs, regardless of actual time taken, accounting for about four hours each.

\section{Results and discussion}
\subsection{Pressure and temperature effects}
\begin{figure}
     \centering
     \begin{subfigure}
         \centering
         \includegraphics[scale=1]{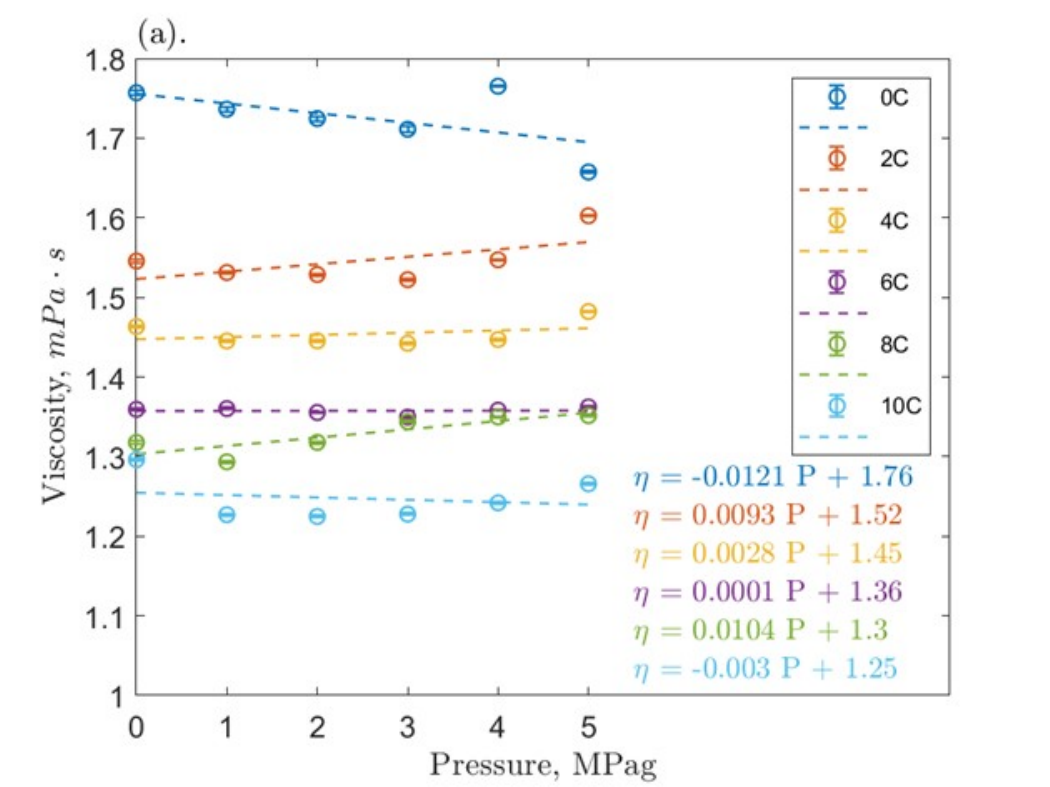}
     \end{subfigure}
     
     \begin{subfigure}
         \centering
         \includegraphics[scale=1]{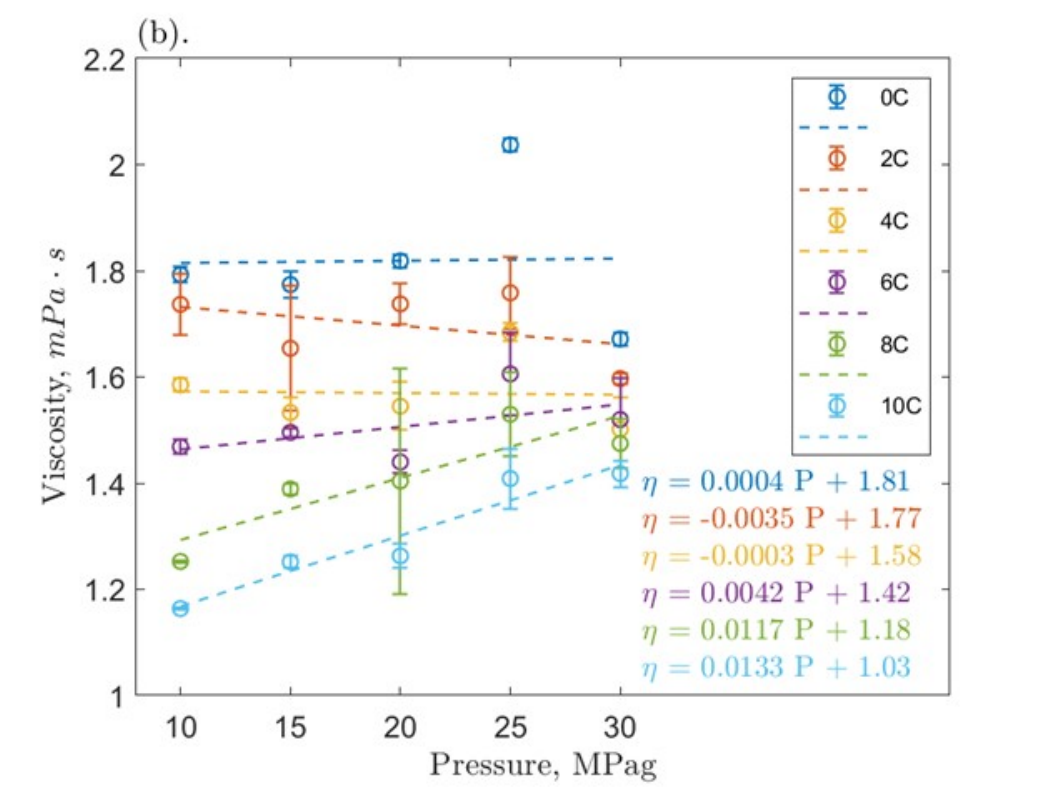}
     \end{subfigure}
     \caption{Pressure effect on the measured viscosity of the methane-water hydrate system studied (a) below 10 MPag and (b) at or above 10 MPag; error bars represent the 95\% confidence intervals on the measured mean viscosity; linear regressions are provided for each condition and presented on the figure.}
     \label{fig:5presseffectch4}
\end{figure}

\begin{figure}
     \centering
     \begin{subfigure}
         \centering
         \includegraphics[scale=1]{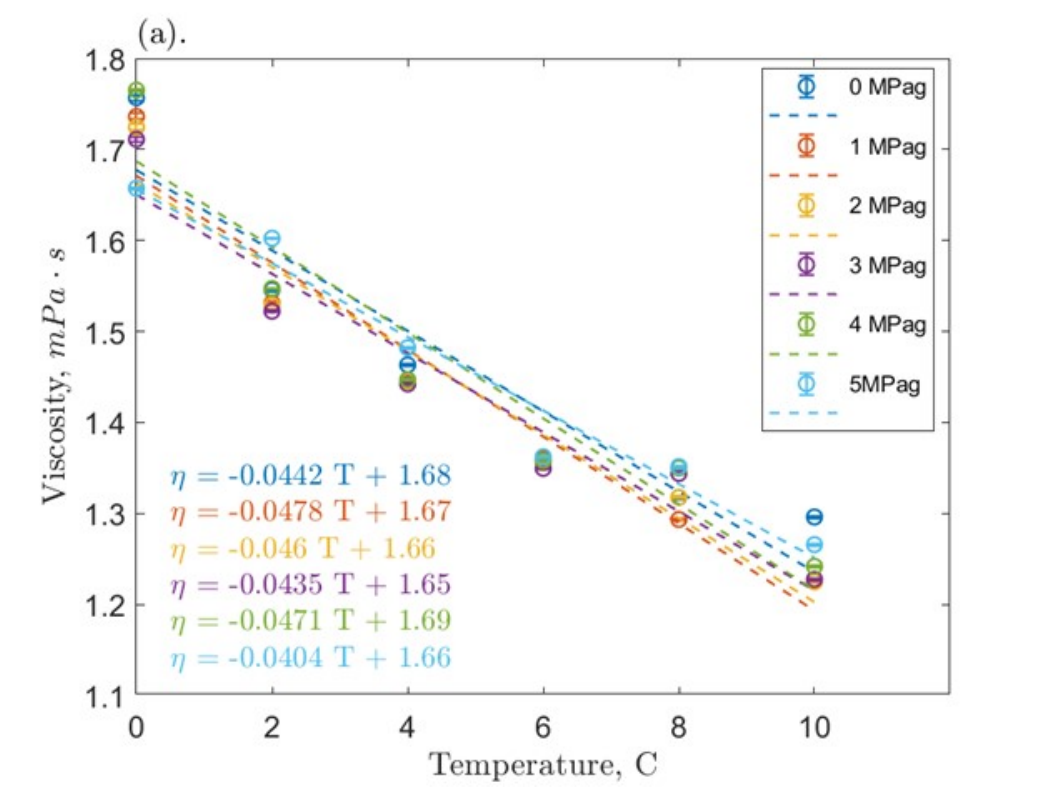}
     \end{subfigure}
     
     \begin{subfigure}
         \centering
         \includegraphics[scale=1]{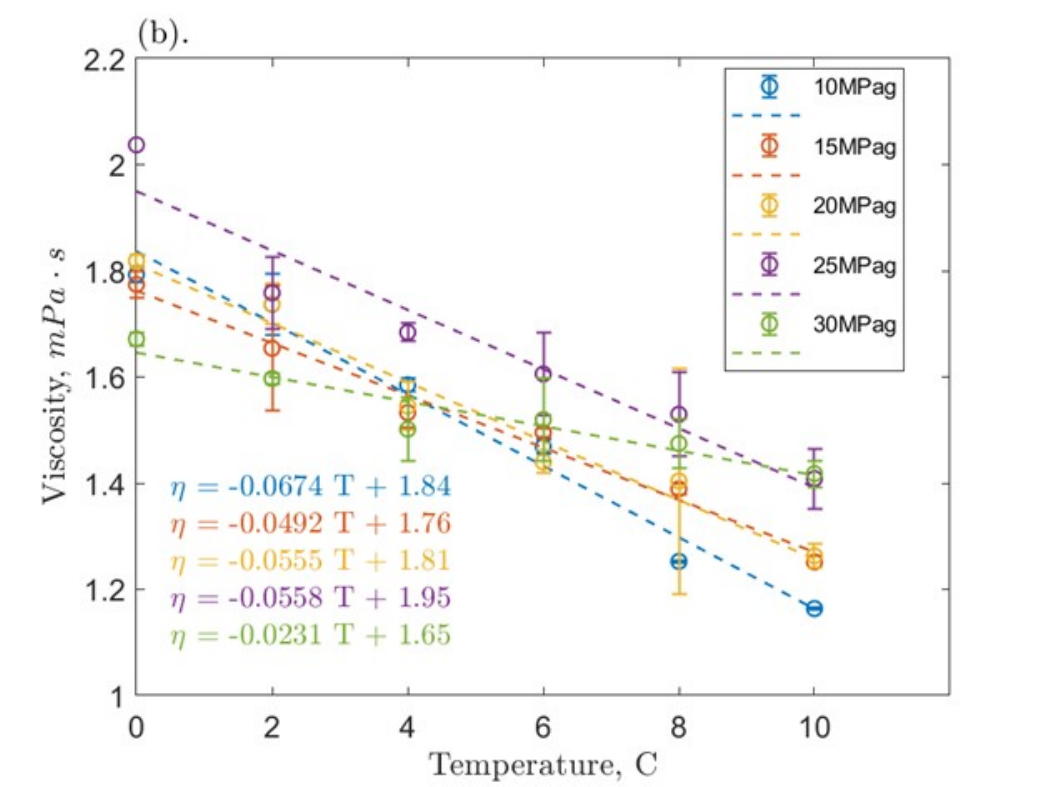}
     \end{subfigure}
     \caption{Temperature effect on the measured viscosity of the methane-water hydrate system studied (a) below 10 MPag and (b) at or above 10 MPag; error bars represent the 95\% confidence interval on the measured mean viscosity; linear regressions are provided for each condition and presented on the figure.}
     \label{fig:6tempeffectch4}
\end{figure}

\begin{figure}
     \centering
     \begin{subfigure}
         \centering
         \includegraphics[scale=1]{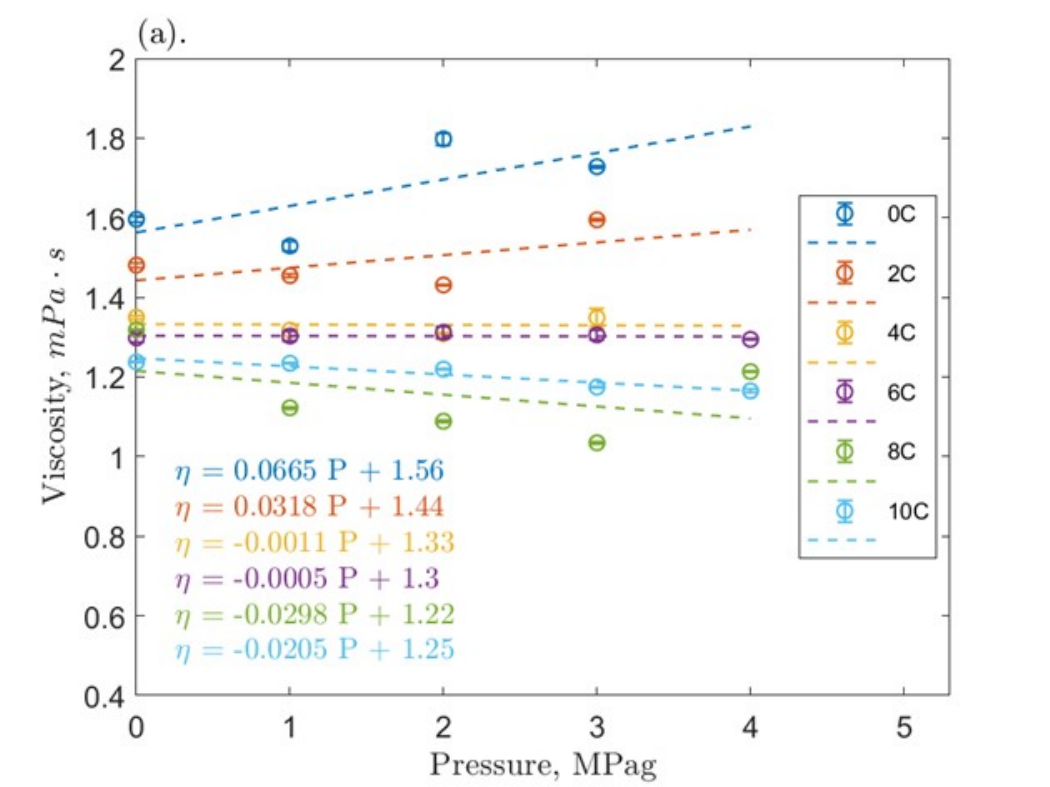}
     \end{subfigure}
     
     \begin{subfigure}
         \centering
         \includegraphics[scale=1]{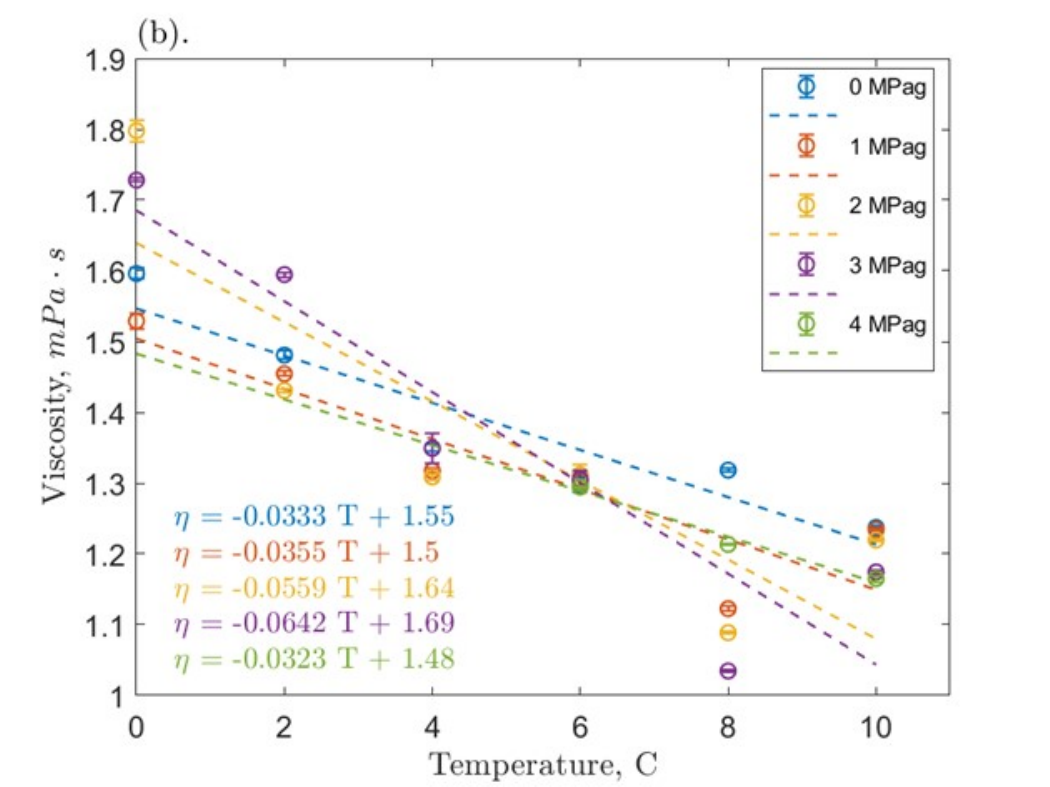}
     \end{subfigure}
     \caption{Pressure effect (a) and temperature effect (b) on the measured viscosity of the carbon dioxide-water systems studied; error bars represent the 95\% confidence interval on the mean viscosity measured; linear regressions are provided for each condition and presented on the figure.}
     \label{fig:7presstempeffectco2}
\end{figure}

Previous rheological studies of hydrate systems have used a relative viscosity definition to facilitate the comparison of results across different types of oil emulsions and for different volume fractions of water content\cite{Majid2018}. This work focuses on pure water systems and thus uses absolute viscosities for rheological characterization of the systems studied. The first pure water system considered was pressurized by methane gas to various pressures and the viscosity was measured isothermally. Part of the objective of this work is to characterize gas hydrate systems before hydrate formation for cross-comparisons in future work involving similar systems with additive species (e.g., polymer inhibitors, nanofluid promoters). Multiple experimental conditions examined in this work involved non-hydrate forming conditions (marked as ‘o’ in Figure~\ref{fig:4expcond}). These results are presented in Figure~\ref{fig:5presseffectch4}, Figure~\ref{fig:6tempeffectch4}, and Figure~\ref{fig:7presstempeffectco2}. The constant shear rate measurement allowed us to take repeated viscosity measurements of the stable (no hydrate formation) system at the same conditions within one test run. The collection of 10-minute periods of stable viscosity measurement (+/- 0.05 mPa$\cdot$s) at an interval of 5.6 seconds resulted in a time average of approximately 110 viscosity data points per test condition presented below. This was done to ensure gas dissolution and temperature fluctuations did not affect the measured viscosity. Linear regressions were implemented to enable relativistic conclusions to be made between the two gas hydrate systems and across experimental conditions explored in this work.

A weak effect of pressure on viscosity was observed ranging from -0.0121 to 0.0104 mPa$\cdot$s/MPag at lower pressures (Figure~\ref{fig:5presseffectch4}a), and at higher pressures ranging from -0.0035 to 0.0133 mPa$\cdot$s/MPag (Figure~\ref{fig:5presseffectch4}b). Additionally, at higher pressures (Figure~\ref{fig:5presseffectch4}b), the effect was increasingly positive as the temperature increased. A negative temperature effect on viscosity is evident from Figure~\ref{fig:6tempeffectch4}, where all pressures considered resulted in viscosity rates of change in the same order of magnitude and ranging from -0.0231 to -0.0674 mPa$\cdot$s/\textcelsius{}. Moreover, the weak pressure effect on viscosity noted above was also evident from the similarity in the parameters of the linear regressions of the isobaric test runs (Figure~\ref{fig:6tempeffectch4}a). Using the 0 MPag test run as a reference, the rates of change of viscosity with respect to temperature across all isobaric test runs varied between -48 to +52\%, while the viscosity at 0\textcelsius{} (y-intercept) varied between -2 to +16\% (Table~\ref{tab:linearparams}). The viscosity measurements in test runs where hydrates were formed (10 to 30 MPag) were more variable as indicated by the larger 95\% confidence intervals (Figure~\ref{fig:5presseffectch4}b and Figure~\ref{fig:6tempeffectch4}b). The variability in the measured viscosities may have been resultant from the nucleation phase in the hydrate formation process (Figure~\ref{fig:1ghkineticphases}) in which hydrate nuclei form and dissociate continuously prior to the onset of the hydrate growth phase. The length of the hydrate nucleation phase is well established to be stochastic at lower pressure driving forces progressively decreasing in stochasticity at higher pressure driving forces\cite{Sloan2008}. Unfortunately, the equipment used in this work was not equipped to measure nucleation times. Additionally, the liquid volume in the systems studied in this work is very small (7.5 mL), and it is exposed to high hydrate forming driving forces (up to 27 MPag above the equilibrium pressure). Under these conditions, very short nucleation times are expected. In other words, even if nucleation times were measurable by the system, the low liquid volume and high driving forces in the experimental conditions investigated would lead to nearly indistinguishable nucleation times between the systems studied.

\begin{table}
\centering
  \caption{Parameters of the linear regressions of the temperature effect on the measured viscosity of the methane-water hydrate system studied and their percentage differences in reference to the 0 MPag test.}
  \label{tab:linearparams}
  \begin{tabular}{c|cccc}
    \hline
    Pressure & Rate of & y-intercept & \% Difference & \% Difference \\
    &  change (m) & (b) & m & b \\
    MPag & mPa$\cdot$s MPag$^{-1}$ & mPa$\cdot$s & - & - \\
    & $\pm$0.0005 & $\pm$0.005 & $\pm$0.5 & $\pm$0.5 \\
    \hline
    0 & -0.044 & 1.68 & - & - \\
    1 & -0.048 & 1.67 & 7 & -1 \\
    2 & -0.046 & 1.66 & 4 & -1 \\
    3 & -0.044 & 1.65 & -2 & -2 \\
    4 & -0.047 & 1.69 & 7 & 1 \\
    5 & -0.040 & 1.66 & -9 & -1 \\
    10 & -0.067 & 1.84 & 52 & 10 \\
    15 & -0.049 & 1.76 & 11 & 5 \\
    20 & -0.056 & 1.81 & 26 & 8 \\
    25 & -0.056 & 1.95 & 26 & 16 \\
    30 & -0.023 & 1.65 & -48 & -2 \\
    \hline
  \end{tabular}
\end{table}

The second gas hydrate system considered was pure water pressurized by carbon dioxide gas. Due to carbon dioxide’s phase envelope, the experimental pressures were limited to 3 and 4 MPag depending on the temperature condition to prevent the liquification of the gas (Figure~\ref{fig:4expcond}b). Figure~\ref{fig:7presstempeffectco2} summarizes the pressure and temperature effects on the system’s measured viscosity. As in the methane hydrate system, the effect of temperature on the viscosity of the carbon dioxide hydrate system is greater than the pressure effect and negative for all isobaric conditions. The temperature effect ranged from -0.0642 to -0.0323 mPa$\cdot$s/\textcelsius{} (Figure~\ref{fig:7presstempeffectco2}b), while the pressure effect ranged from -0.0298 to 0.0665 mPa$\cdot$s/MPag. Additionally, the linear regressions for the carbon dioxide system demonstrate changes in viscosity with pressure up to an order of magnitude larger than in the methane system at similar pressures (Figure~\ref{fig:5presseffectch4}a and Figure~\ref{fig:7presstempeffectco2}a). This was likely due to a higher amount of carbon dioxide present in water, as its solubility in water has been shown to be an order of magnitude greater than that of methane\cite{Servio2001,Servio2002}. The findings above suggest that temperature changes may be a more significant parameter than pressure changes when considering viscosity in the design of new methane and carbon dioxide hydrate technologies at extreme high pressures.

\begin{figure}
     \centering
     \begin{subfigure}
         \centering
         \includegraphics[scale=1]{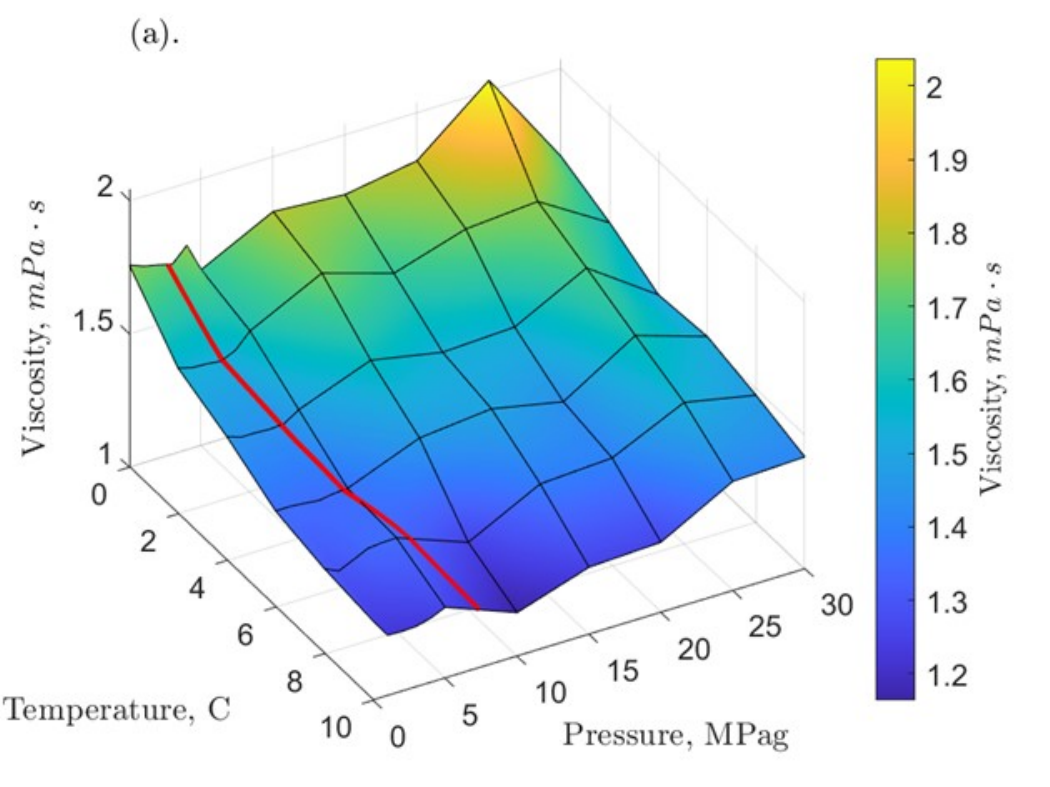}
     \end{subfigure}
     
     \begin{subfigure}
         \centering
         \includegraphics[scale=1]{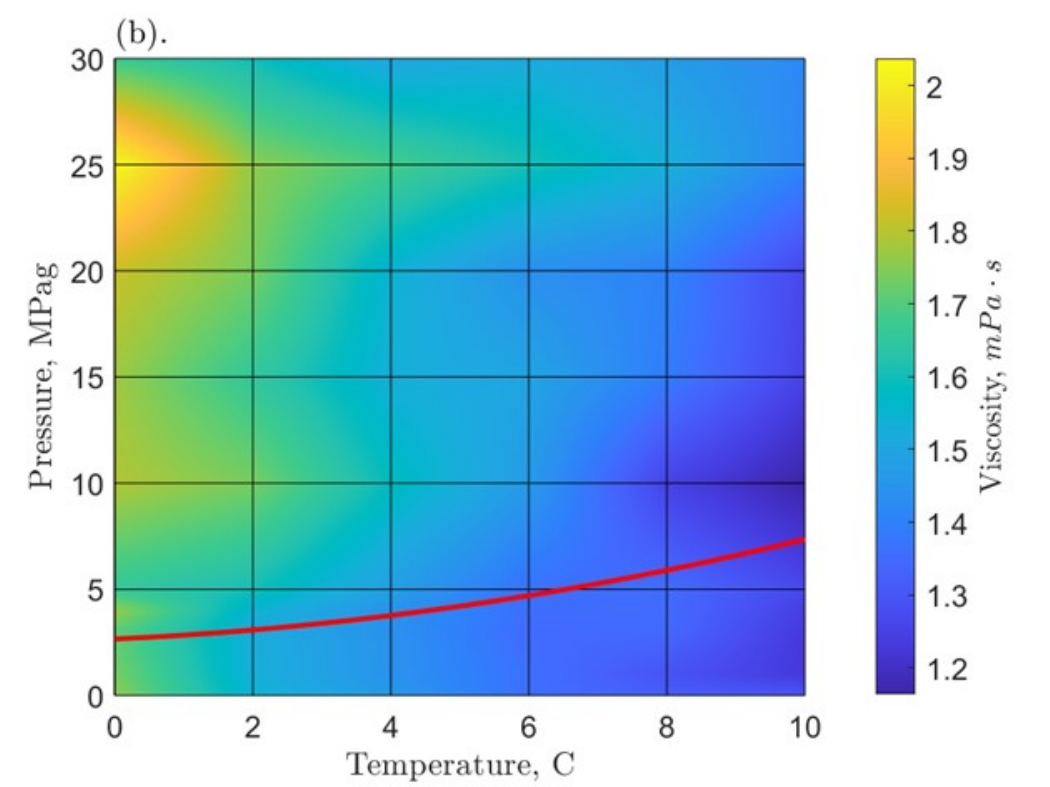}
     \end{subfigure}
     \caption{Rheological phase diagram of the methane hydrate system; (a) interpolated surface of the measure viscosity data; (b) planar view of the rheological phase diagram; the diagram contains isobaric and isothermal lines in black and the H-L-V equilibrium line of methane hydrates in red.}
     \label{fig:8rheophasediagch4}
\end{figure}

\begin{figure}
     \centering
     \begin{subfigure}
         \centering
         \includegraphics[scale=1]{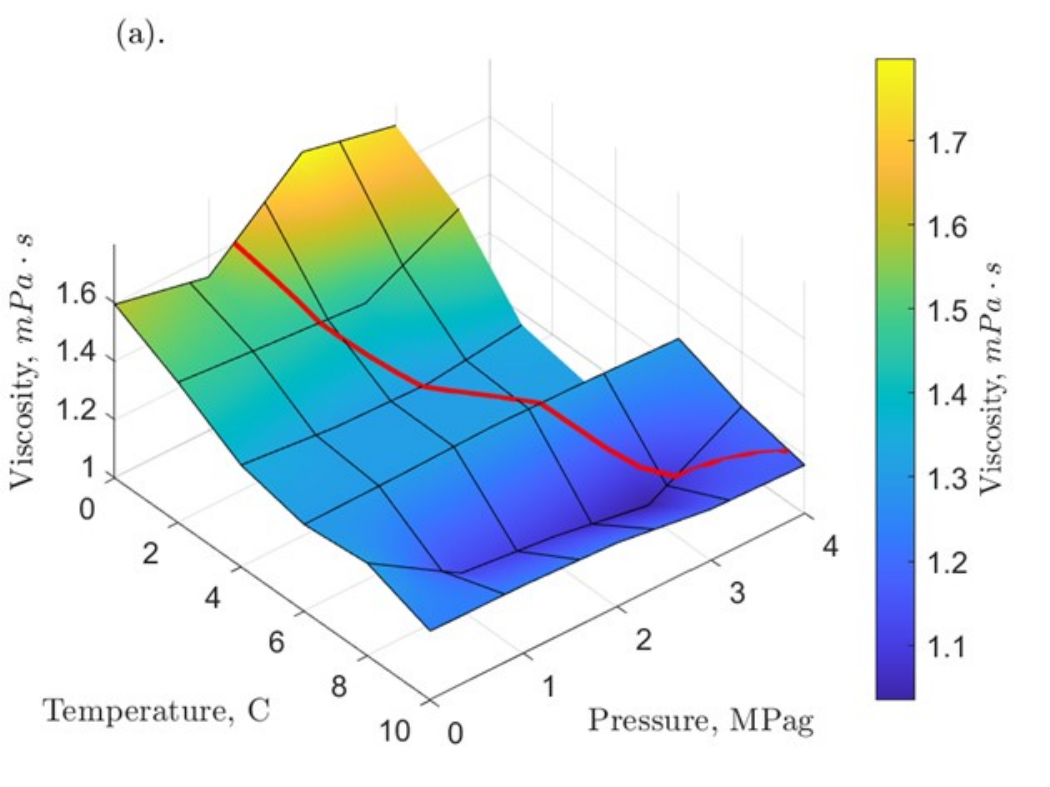}
     \end{subfigure}
     
     \begin{subfigure}
         \centering
         \includegraphics[scale=1]{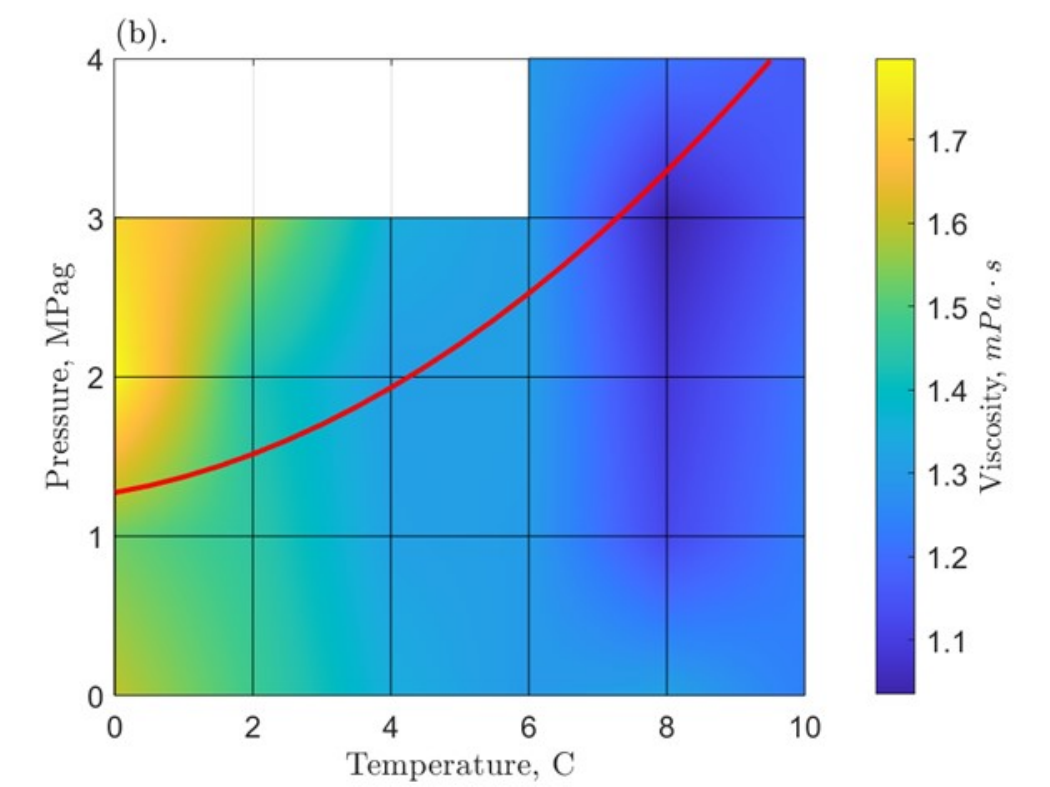}
     \end{subfigure}
     \caption{Rheological phase diagram of the carbon dioxide hydrate system; (a) interpolated surface of the measure viscosity data; (b) planar view of the rheological phase diagram; the diagram contains isobaric and isothermal lines in black and the H-L-V equilibrium line of carbon dioxide hydrates in red.}
     \label{fig:9rheophasediagco2}
\end{figure}

This work allowed for the characterization of the rheology of the methane and carbon dioxide hydrate systems throughout the regions of the thermodynamic phase diagram (Figure~\ref{fig:4expcond}) to extend it to extreme high pressures. A rheological phase diagram was developed for the methane hydrate system presented above. Figure~\ref{fig:8rheophasediagch4} presents an interpolated viscosity surface for the range of experimental conditions of the methane system. It also includes an overlaid H-L-V equilibrium line to distinguish the H-L and L-V phase regions above and below the equilibrium line, respectively (as presented in Figure~\ref{fig:4expcond}a). Figure~\ref{fig:9rheophasediagco2} is the rheological phase diagram for the carbon dioxide hydrate system. A rheological phase diagram such as these may be useful tools for control, prediction, and design of gas hydrate technologies.

The systems studied here are composed mostly of pure water. The viscosity of water has been previously studied and observed to behave as a non-simple fluid\cite{Bett1965,Bridgman1925,Gallo2016,Horne1966,Wonnham1967}. The non-simple fluid behaviour is attributed to an observed initial decrease in viscosity with increasing pressure followed by an approximately linear increase leading to a local minimum in viscosity. This minimum has been shown to occur around 150-200 MPag and to be more pronounced at lower temperatures (near 2\textcelsius{})\cite{Horne1966}. The high pressures achieved in this work do not come close to the expected pressure range to observe the viscosity minimum. However, a linear decrease in viscosity was observed in the methane system at lower temperatures for the pressures explored, (Figure~\ref{fig:5presseffectch4}) indicating a possible minimum at higher pressures. Conversely, an increase in viscosity was observed in the carbon dioxide system at lower temperatures (Figure~\ref{fig:7presstempeffectco2}). Due to physical challenges to achieve higher experimental pressures, the results of this work suggest that computational methods such as molecular dynamics be used to calculate the viscosity of the methane systems above 30 MPag to further explore the possibility of a viscosity minimum in methane-water systems near hydrate forming conditions.

\subsection{Liquid-to-solid phase transition}
\begin{figure}[ht]
\centering
\includegraphics[scale = 1]{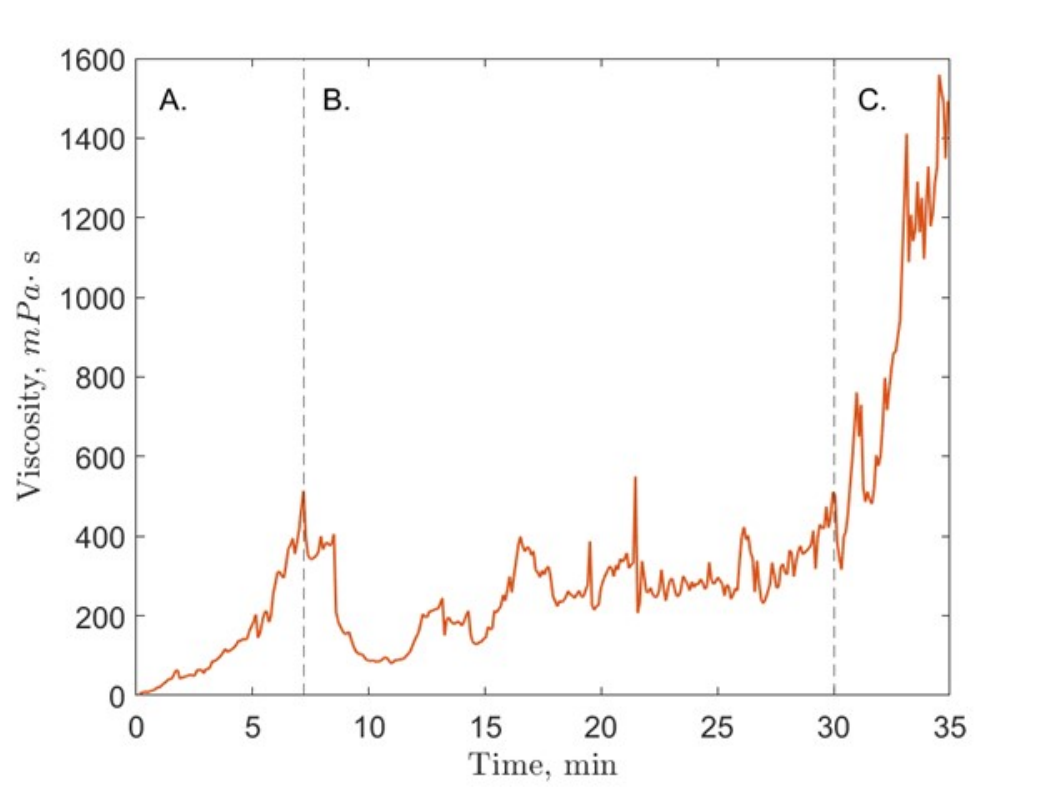}
\caption{Example of the three stages in hydrate growth as measured by viscosity in the carbon dioxide test run at 2\textcelsius{} and 30 MPag; A: initial growth, B: slurry phase, and C: final growth.}
\label{fig:10viscstages}
\end{figure}

\begin{figure}[ht]
\centering
\includegraphics[scale = 0.8]{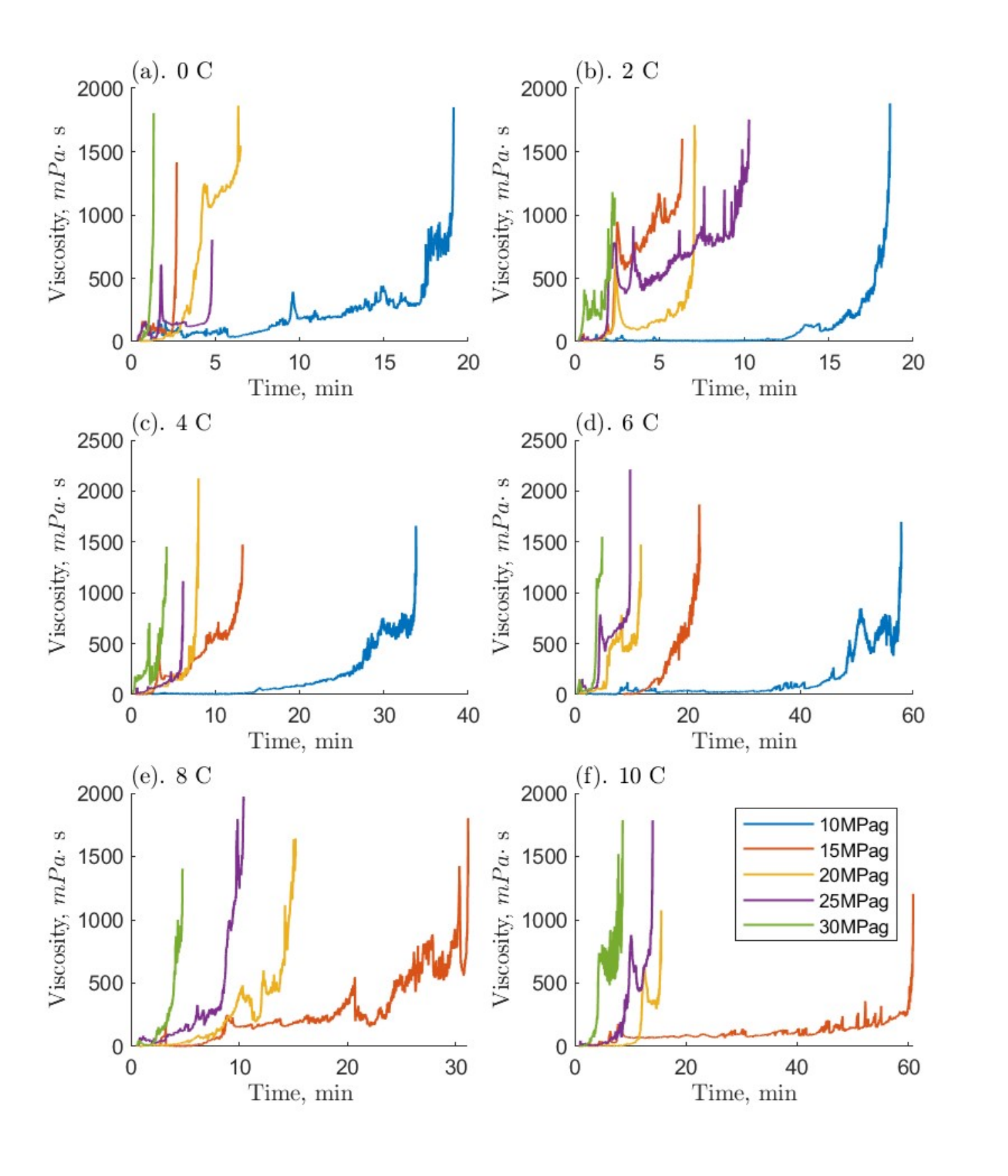}
\caption{Measured viscosity of methane-water systems where hydrate formation occurred; each sub-panel (a-f) separates test runs by temperature and contains isobaric viscosity time series starting at onset of hydrate formation.}
\label{fig:11viscpresseffect}
\end{figure}

The presence of a phase transition was evident from the temporal viscosity data collected. The phase transition was characterized here by an unstable viscosity that fluctuated and increased over time until a maximum was measured. Figure~\ref{fig:10viscstages} presents the carbon dioxide test run at 2\textcelsius{} and 3 MPag, where the transition through the slurry phase was clear and can be used for a general description of the phase transition process. Hydrate growth is initially detected by increased measured viscosity and occurs unencumbered in zone (A) in Figure~\ref{fig:10viscstages}. At a certain point, the hydrate growth is disturbed and does not proceed at the same rate as previously. In zone (B), the hydrate growth seems to be limited; the viscosity changes in a nonuniform manner and does not progress upwards significantly. Over time, however, the viscosity increases, and it reaches a point where hydrate growth rate seems to increase again as the viscosity dramatically rises and ends at a maximum value (Figure~\ref{fig:10viscstages}C). The system transitions from the liquid to the solid phase in a slurry – a suspension of hydrate clusters in the liquid phase. The slurry is formed at the onset of the hydrate growth phase (Figure~\ref{fig:1ghkineticphases}). This general process was observed in all hydrate forming test runs in this work. Test runs with progressively larger driving forces had shorter slurry phases in the hydrate growth process. In certain cases, the driving force was so large that the slurry phase was not detected by the viscosity measurements (Figure~\ref{fig:11viscpresseffect}). 

The viscosity behaviour of the test runs with successful methane hydrate formation are presented in Figure~\ref{fig:11viscpresseffect}. These figures demonstrate how increments in pressure and temperature, respectively, change the viscosity behaviour during the hydrate growth phase. Moreover, the temperature and pressure effects on the slurry phase length was evident. Hydrate formation occurred more rapidly at higher pressures and lower temperatures (higher driving forces) and the slurry phase was short enough to seem absent. In Figure~\ref{fig:11viscpresseffect}, the shortest phase transition stages were observed at 0 and 2\textcelsius{} The phase transition length was found to increase for test runs with decreasing driving forces (higher temperatures and lower pressures). Figure~\ref{fig:11viscpresseffect}f demonstrates the considerably longer slurry phase transition for the 15 MPag test condition compared to test runs at higher pressures.

\begin{figure}
     \centering
     \begin{subfigure}
         \centering
         \includegraphics[scale=1]{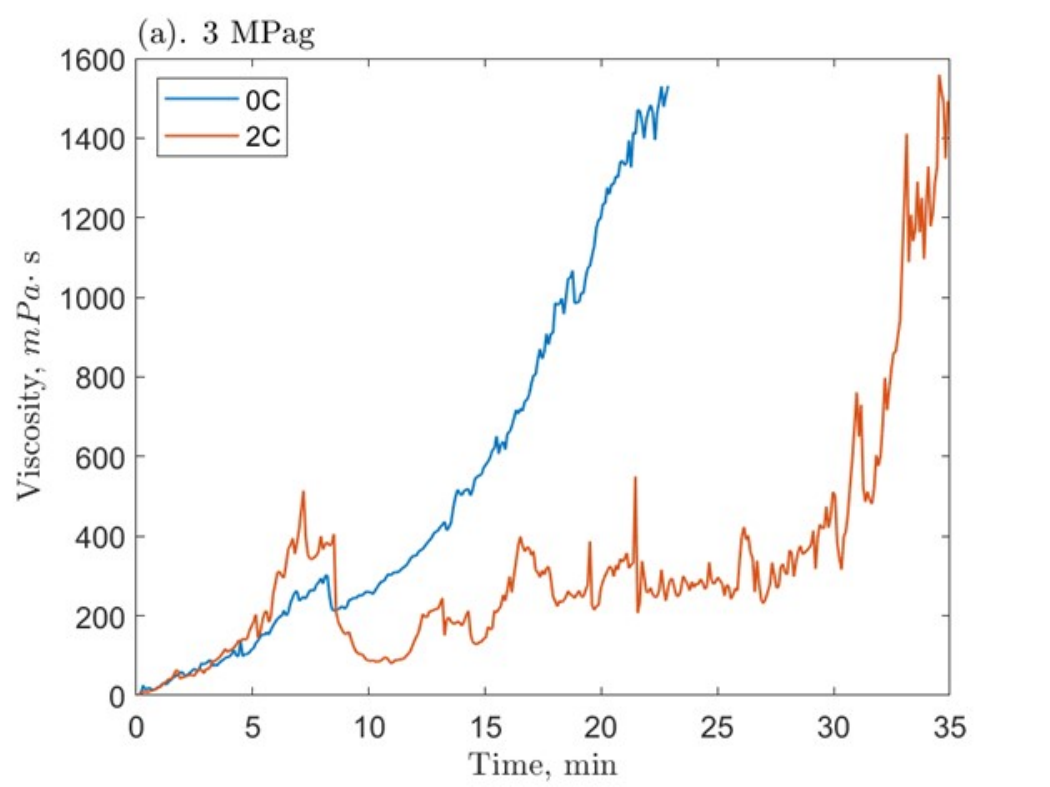}
     \end{subfigure}
     
     \begin{subfigure}
         \centering
         \includegraphics[scale=1]{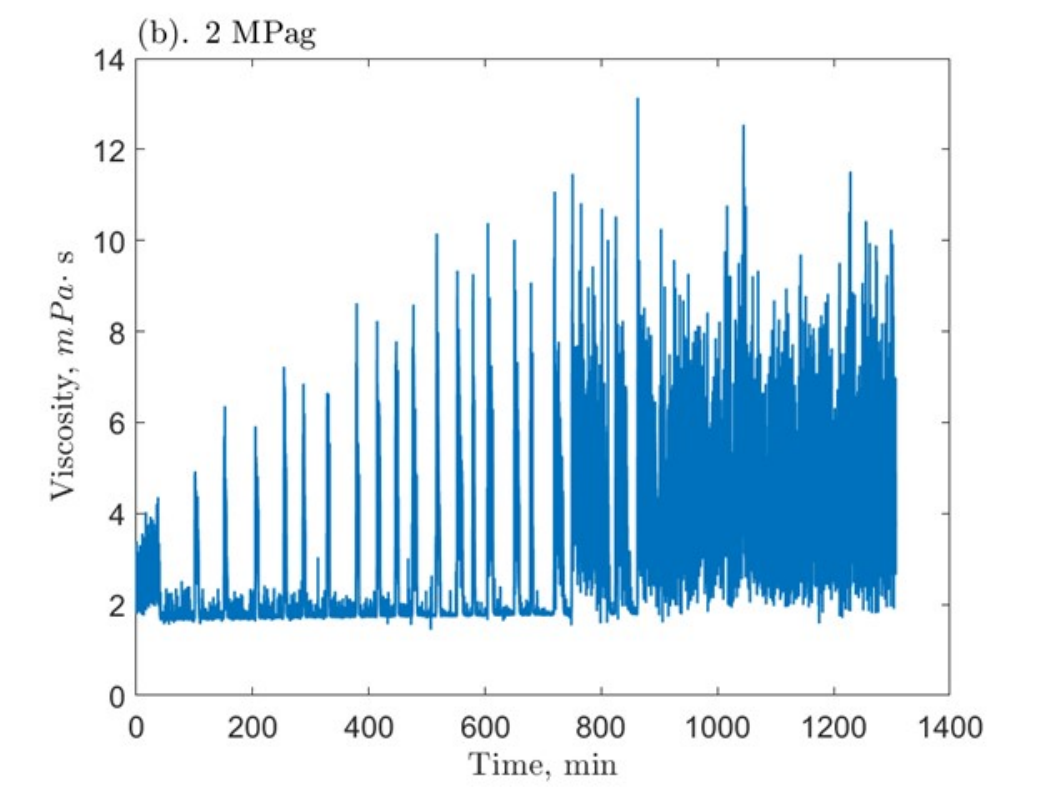}
     \end{subfigure}
     \caption{Measured viscosity of carbon dioxide-water systems where hydrate formation occurred; each sub-panel separates test runs by pressure and contains isothermal viscosity time series starting at onset of hydrate formation.}
     \label{fig:12co2viscinstability}
\end{figure}

The phase transition was also observed in the test runs where successful carbon dioxide hydrates were formed. Figure~\ref{fig:12co2viscinstability} presents the viscosity profile of the three carbon dioxide hydrate test runs. Figure~\ref{fig:12co2viscinstability}a, contrasts the viscosity behaviour of the system for conditions where a fast (short) phase transition stage (at 0\textcelsius{}) and a slow (long) phase transition stage (at 2\textcelsius{}) occurred at 3 MPag. As for the methane system, the long phase transition stage is characterized by an unstable viscosity progression up to its maximum value. Figure~\ref{fig:12co2viscinstability}b further demonstrates the instability in viscosity during hydrate formation. In this case, however, the hydrate formation did not reach a maximum value; the system fluctuated between its baseline (initial) viscosity and values up to one order of magnitude higher (Table~\ref{tab:2co2maxvisc}), but hydrate growth did not progress enough in the 24-hour period to achieve maximum viscosity. This test run had one of the lowest driving forces that was still considered hydrate forming by this work (726 kPag). The inability of this system condition to sustain hydrate growth confirms the assumption that driving forces below 50 kPag would be too low to form hydrates within a reasonable time length. The instability of viscosity measurements in the phase transition and the inability of low driving forces to sustain hydrate growth can be attributed to system limitations. These limitations will be discussed in the next section.

\begin{table}
\centering
  \caption{Time required for carbon dioxide-water system to reach maximum viscosity from the onset of hydrate formation; driving force includes calculation uncertainties, while max viscosity and time to max viscosity include uncertainties associated with significant figures of measurements.}
  \label{tab:2co2maxvisc}
  \begin{tabular}{c|c|ccc}
    \hline
    Temperature & Pressure & Driving force & Max viscosity & Max viscosity \\
    \textcelsius{} & MPag & MPag & mPa$\cdot$s & minutes \\
    \hline
    0 & 2 & 0.73$\pm$0.005 & 13.0$\pm$0.05 & 930$\pm$0.5 \\
    & 3 & 1.7$\pm$0.05 & 1530$\pm$5 & 23$\pm$0.5 \\
    \hline
    2 & 3 & 1.5$\pm$0.05 & 1490$\pm$5 & 35$\pm$0.5 \\
    \hline
  \end{tabular}
\end{table}

A decomposition of the data from unstable carbon dioxide test run was performed as a further way to characterize the pre-transitional phenomena occurring in Figure~\ref{fig:12co2viscinstability}b. The analysis was performed over the first 850 minutes of the test run, where spikes in viscosity were most clearly demarcated. The region past this point was too unstable to completely attribute the fluctuation to the pre-transitional phase phenomena; ball bearing wear may also have been a contributing factor to fluctuations in viscosity in the late portion of this test run. The length of each viscosity spike and the magnitude of the spikes are presented in Table~\ref{tab:3co2viscinstability}. The nineteen cycles of viscosity spikes identified had an average frequency of 3.7x10$^{-4}$ Hz (approximately 45 minutes per cycle) over the period analyzed and an average viscosity maximum magnitude of 8.7 mPa$\cdot$s.

\begin{table}
\centering
  \caption{Spikes in measured viscosity for the carbon dioxide test run at 0\textcelsius{} and 2 MPag.}
  \label{tab:3co2viscinstability}
  \begin{tabular}{c|cc|cc}
    \hline
    Cycles & Start time & End time & Max viscosity & Duration of spike \\
    & minutes & minutes & mPa$\cdot$s & minutes \\
    & $\pm$0.05 & $\pm$0.05 & $\pm$0.05 & $\pm$0.05 \\
    \hline
    1 & 100.1 & 108.6 & 4.9 & 8.5 \\
    2 & 150.1 & 158.2 & 6.4 & 8.1 \\
    3 & 204.2 & 211.5 & 5.9 & 7.3 \\
    4 & 252.6 & 259.1 & 7.2 & 6.5 \\
    5 & 285.4 & 291.1 & 6.9 & 5.7 \\
    6 & 327.0 & 333.2 & 6.7 & 6.2 \\
    7 & 377.2 & 384.3 & 8.6 & 7.2 \\
    8 & 413.0 & 421.1 & 8.2 & 8.1 \\
    9 & 444.5 & 452.0 & 7.8 & 7.5 \\
    10 & 474.1 & 483.7 & 8.6 & 9.6 \\
    11 & 515.1 & 524.9 & 10.1 & 9.8 \\
    12 & 550.8 & 563.1 & 9.3 & 12.3 \\
    13 & 577.7 & 584.7 & 9.3 & 7.0 \\
    14 & 603.3 & 615.8 & 10.4 & 12.5 \\
    15 & 648.9 & 661.3 & 10.0 & 12.3 \\
    16 & 676.9 & 683.0 & 9.1 & 6.2 \\
    17 & 717.4 & 736.3 & 11.1 & 18.9 \\
    18 & 747.9 & 816.1 & 11.5 & 68.2 \\
    19 & 823.4 & 854.5 & 13.1 & 31.1 \\
    \hline
  \end{tabular}
\end{table}

\begin{figure}
     \centering
     \begin{subfigure}
         \centering
         \includegraphics[scale=1]{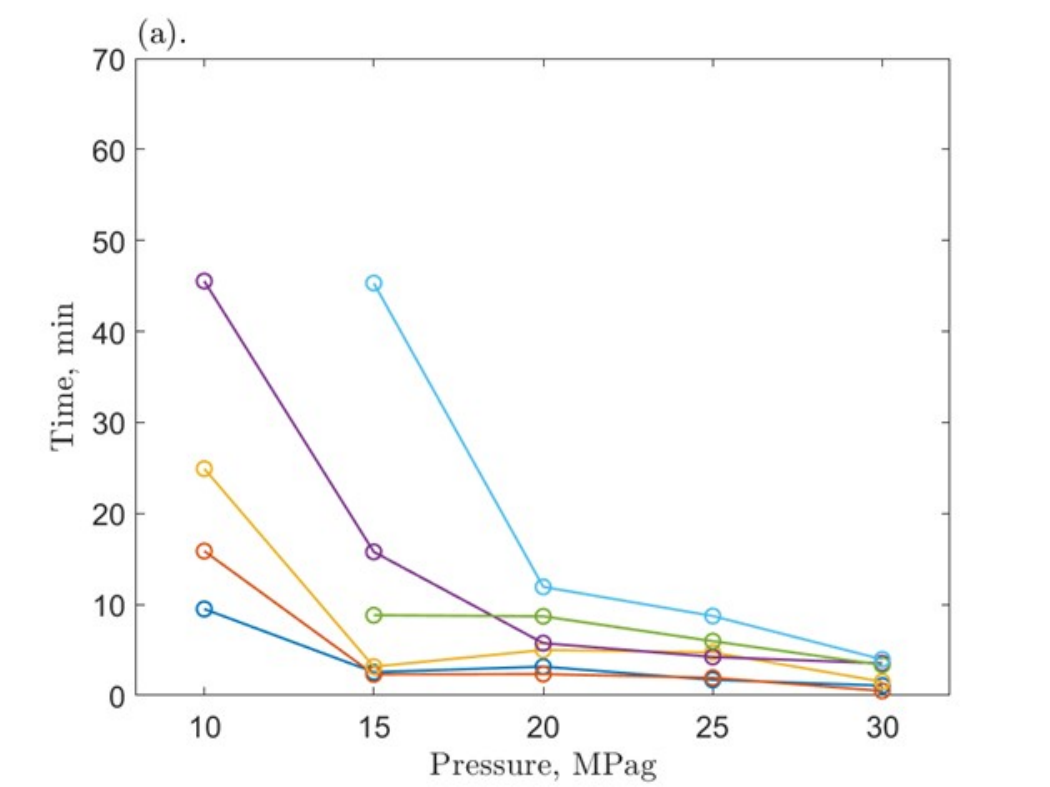}
     \end{subfigure}
     
     \begin{subfigure}
         \centering
         \includegraphics[scale=1]{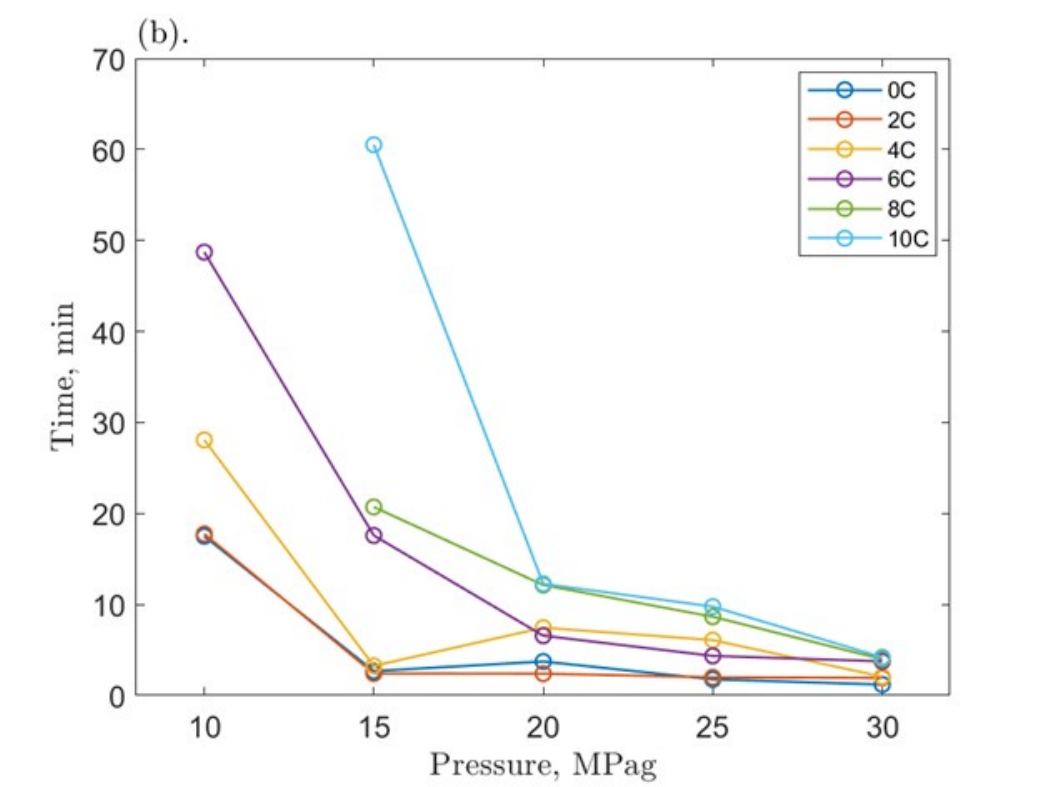}
     \end{subfigure}
     \caption{Time required for methane-water system to reach (a) 200 mPa$\cdot$s and (b) 500 mPa$\cdot$s from the onset of hydrate formation at various pressures.}
     \label{fig:13timetovisc}
\end{figure}

The length of the phase transition affects the time required for the system to reach maximum viscosity. Figure~\ref{fig:13timetovisc} presents the pressure effect on the time required for the methane hydrate system to achieve representative viscosities (200 and 500 mPa$\cdot$s), and thus gives insight into to the length of the phase transition stage for each test run. Additionally, these times may be valuable for process and equipment design considerations involving similar hydrate systems. The time required for the system to reach viscosities higher than 500 mPa$\cdot$s was negligibly higher, and thus were omitted here. The greatest decrease in time required for the system to achieve these viscosities was observed between 10 and 15 MPag (Figure~\ref{fig:13timetovisc}); the time required was approximately halved for all temperatures observed. The carbon dioxide hydrate system did not produce as many successful tests in which hydrates were formed for the same analysis. However, Table~\ref{tab:2co2maxvisc} presents the maximum viscosity, the time required to achieve it, and the driving forces for the two test runs where hydrates were formed (Figure~\ref{fig:12co2viscinstability}a), and for the test run where hydrates began to form but did not achieve maximum viscosity (Figure~\ref{fig:12co2viscinstability}b). It is noteworthy that the maximum viscosity reported is solely an indication of the reading that occurred when the rheometer reached its torque limit (150 mN$\cdot$m). Repeated measurements would likely result in different values in the same order of magnitude. However, the time required to reach these values are expected to be dependent on the driving force available and thus to be similar if repeated. The 0\textcelsius{} and 2 MPag test run only had a 726 kPag driving force and it exhausted the maximum 24-hour period for hydrate formation. Table~\ref{tab:2co2maxvisc} lists its instantaneous maximum measured viscosity within the 24-hour period, but it is not associated with the hydrate solidification normally associated with the maximum viscosities reported for successful hydrate formation in this work.

\subsection{System limitations to hydrate formation}
The study of rheological properties in gas hydrate systems requires the use of specialized rheometers with high pressure ratings to achieve the necessary conditions for hydrate formation. This poses a challenge for this type of research. This work used a rheometer rated to 40 MPag to reach experimental pressures of 30 MPag. However, this work observed conditions with considerable driving forces for hydrate formation fail to form hydrates in the allowed 24-hour period. This indicates limitations which may be inherent to the measurement system. The failed driving forces were as high as 4.1 MPag for methane and 2.07 MPag for carbon dioxide. In Figure~\ref{fig:4expcond}, blue ‘x’s specify the experimental conditions where hydrates formation was observed along with their driving forces. While black ‘x’s are the conditions above the H-L-V equilibrium line in which hydrate formation is thermodynamically favourable, but where no hydrate formation was observed in this work. As in Figure~\ref{fig:4expcond}, the red ‘x’s mark conditions which were considered to have driving forces too low for hydrate formation for the purpose of this work.

Some of the possible system limitations that may be responsible for failed driving forces are postulated to be kinetic, mass diffusion, and heat of crystallization effects. First, limitations on the nucleation of hydrates cause a kinetic impediment to hydrate formation. The rheometer system is, by design, a high-shear environment which may cause the mechanical dissociation of newly formed gas hydrate nuclei during the nucleation phase. The vorticity of the shear flow lacks directionality, and the orientation of extensional effects are $\pm$45 degrees to flow direction. Together, they tend to separate neighbouring groups of fluid components, thereby dissociating hydrate clusters. The nuclei need to reach a critical radius before the growth phase of hydrate formation can begin\cite{Sloan2008}. If the mechanical dissociation prevents the agglomeration of nuclei or directly causes the dissociation of nuclei, this can reduce the effective driving force to hydrate formation. Additionally, the rheometer measurement system and sample well are smooth stainless-steel surfaces that do not provide many sites for hydrate nucleation. The lack of nucleation sites is exacerbated by the ultra-pure RO water used, which contains few impurities that could provide nucleation sites. Second, the volume and geometry of the system introduce a diffusion limitation to hydrate formation. The rheometer sample well has a small volume and due to its double annulus geometry (Figure~\ref{fig:3rheometer}B) a low gas-liquid surface area is available for mass diffusion, which limits the dissolution of gas into the liquid phase. Finally, the hydrate formation process is exothermic, which may under certain circumstances be thermodynamically self-limiting\cite{Sloan2008}. If the system conditions are favourable, the change in temperature may not interfere significantly with the start of the growth phase in hydrate formation. However, in conjunction with the kinetic and diffusion limitations suggested above, the heat of hydrate formation may pose a heat effect limitation. The change in temperature during hydrate formation may contribute to inhibit the agglomeration of hydrate nuclei and to prevent reaching critical mass. Heat effect in carbon dioxide hydrate formation has been shown to be high enough to limit the effect of certain nanofluid promoters\cite{Pasieka2015}. 

The kinetic, diffusion, and heat of crystallization effect limitations suggested above were also evident during successful hydrate formation. The intermediate pressure conditions in Figures~\ref{fig:11viscpresseffect} and~\ref{fig:12co2viscinstability} show prolonged slurry phase periods, where the proposed limitations above may prevent the hydrate formation process to progress to the growth phase. In particular, the heat effect limitation would be more pronounced in the carbon dioxide system, as it has a higher heat of formation than methane hydrates\cite{Sloan2008}. The clearest example of this is the carbon dioxide test run at 0℃ and 2 MPag (Figure ~\ref{fig:12co2viscinstability}b). The system periodically increases in viscosity from the initial viscosity of 1.7 mPa$\cdot$s to register a maximum value at 13.1 mPa$\cdot$s. However, the system is incapable of sustaining the hydrate formation that causes the increased viscosity. As a result, its measured viscosity continuously fluctuates as the system tries to reach a balance between hydrate formation and dissociation. Moreover, the methane hydrate test runs also demonstrated self-limiting behaviour. In Figures~\ref{fig:11viscpresseffect}a and~\ref{fig:11viscpresseffect}b, the 20 and 25 MPag test runs had longer slurry phases, in which the viscosity increased to register high values before retracing back to lower values. The driving forces in these cases were high enough, however, that any limitations to hydrate growth were eventually overcome leading to the ultimate maximum viscosity value to be registered.

To mitigate the impact of the effects discussed above, all hydrate forming conditions were given a 24-hour period for the formation of hydrates to occur. This is considerably longer than the time required by all successful driving force conditions to form hydrates for both methane and carbon dioxide (Figure~\ref{fig:13timetovisc} and Table~\ref{tab:2co2maxvisc}). Despite this measure, multiple driving forces failed to form hydrates for both methane and carbon dioxide systems (Figure~\ref{fig:4expcond}). Recently, the effect of mixing by nanofluids in gas hydrate systems has been shown to improve the kinetics of hydrate formation, which could further mitigate these limitations\cite{McElligott2019,McElligott2021a}. However, this would fundamentally change the system, and thus would not have been an adequate consideration in this study. It is suggested, however, for future work to explore the effect of nanofluids on the gas hydrate system viscosities and whether improved mixing can overcome some of the limitations identified here.

\section{Conclusions}
This work explored the extreme high-pressure rheology of methane and carbon dioxide hydrate systems in pure water at various temperatures. The effect of temperature on the viscosity of methane hydrate systems was found to be one order of magnitude larger than the pressure effect for most pressures considered, especially at lower temperatures. The pressure effect on the viscosity of carbon dioxide hydrate systems was one order of magnitude larger than that of the methane hydrate system, likely as there was more gas present in solution due to carbon dioxide’s higher solubility in water. Additionally, linear regressions of the pressure and temperature effects on viscosity were presented for the conditions explored. These results suggest that temperature is an important parameter for controlling viscosity in new hydrate technologies involving gas hydrates from pure water. A novel rheological phase diagram for methane and carbon dioxide hydrates was developed to further characterize these systems. Additionally, a phase transition stage was identified as characterized by the behaviour of the system viscosity as the hydrate formation occurred. The pressure and temperature effects and the time length of the phase transitional stage were also discussed. The time to reach maximum viscosity in methane hydrate systems was found to decrease by at least a factor of two as pressure is increased from 10 to 15 MPag.

Several test conditions with considerable hydrate thermodynamic driving forces were observed to fail in the formation of hydrates. These were as high as 4.1 MPag and 2.07 MPag for the methane and carbon dioxide hydrate systems, respectively. This work considered kinetic, mass diffusion, and heat of crystallization effects limitations to the formation of methane and carbon dioxide hydrates in the extreme high-pressure rheometer used. Due to the physical limitations on the experimental equipment necessary to achieve higher pressures in a rheometer, computational methods such as molecular dynamics were recommended for the calculation of the viscosity of methane and carbon dioxide hydrate systems at pressures above 30 MPag. Future experimental work will use the results presented here as a baseline for comparison to examine the effect of promoter and inhibitor additives on the extreme high-pressure rheology of gas hydrate systems.

\bibliographystyle{abbrvnat}
\bibliography{main.bib}

\begin{thebibliography}{43}
\providecommand{\natexlab}[1]{#1}
\providecommand{\url}[1]{\texttt{#1}}
\expandafter\ifx\csname urlstyle\endcsname\relax
  \providecommand{\doi}[1]{doi: #1}\else
  \providecommand{\doi}{doi: \begingroup \urlstyle{rm}\Url}\fi

\bibitem[Aaron and Tsouris(2005)]{Aaron2005}
D.~Aaron and C.~Tsouris.
\newblock Separation of co2 from flue gas: A review.
\newblock \emph{Separation Science and Technology}, 40:\penalty0 321--348, 1
  2005.
\newblock ISSN 0149-6395.
\newblock \doi{10.1081/SS-200042244}.
\newblock URL \url{https://doi.org/10.1081/SS-200042244}.
\newblock doi: 10.1081/SS-200042244.

\bibitem[Bett and Cappi(1965)]{Bett1965}
K.~E. Bett and J.~B. Cappi.
\newblock Effect of pressure on the viscosity of water.
\newblock \emph{Nature}, 207:\penalty0 620--621, 1965.
\newblock \doi{10.1038/207620a0}.
\newblock URL \url{https://doi.org/10.1038/207620a0}.

\bibitem[Bridgman(1925)]{Bridgman1925}
P.~W. Bridgman.
\newblock The viscosity of liquids under pressure.
\newblock \emph{Proceedings of the National Academy of Sciences of the United
  States of America}, 11:\penalty0 603--606, 1925.
\newblock ISSN 00278424.
\newblock URL \url{www.jstor.org/stable/84781}.

\bibitem[Camargo and Palermo(2002)]{Camargo2002}
R.~Camargo and T.~Palermo.
\newblock Rheological properties of hydrate suspensions in an asphaltenic crude
  oil.
\newblock \emph{4th International Conference on Gas Hydrates}, pages 19--23,
  2002.

\bibitem[Camargo et~al.(2000)Camargo, Palermo, Sinquin, and
  Glenat]{Camargo2000a}
R.~Camargo, T.~Palermo, A.~Sinquin, and P.~Glenat.
\newblock Rheological characterization of hydrate suspensions in oil dominated
  systems.
\newblock \emph{Annals of the New York Academy of Sciences}, 912:\penalty0
  906--916, 1 2000.
\newblock \doi{10.1111/j.1749-6632.2000.tb06844.x}.
\newblock URL \url{https://doi.org/10.1111/j.1749-6632.2000.tb06844.x}.
\newblock doi: 10.1111/j.1749-6632.2000.tb06844.x.

\bibitem[Carroll(2014)]{Carroll2014}
J.~Carroll.
\newblock \emph{Natural Gas Hydrates: A Guide for Engineers}.
\newblock Gulf Professional Publishing, 3rd edition, 2014.

\bibitem[Clain et~al.(2012)Clain, Delahaye, Fournaison, Mayoufi, Dalmazzone,
  and Fürst]{Clain2012}
P.~Clain, A.~Delahaye, L.~Fournaison, N.~Mayoufi, D.~Dalmazzone, and W.~Fürst.
\newblock Rheological properties of tetra-n-butylphosphonium bromide hydrate
  slurry flow.
\newblock \emph{Chemical Engineering Journal}, 193-194:\penalty0 112--122,
  2012.
\newblock ISSN 1385-8947.
\newblock \doi{https://doi.org/10.1016/j.cej.2012.04.027}.
\newblock URL
  \url{http://www.sciencedirect.com/science/article/pii/S1385894712004743}.

\bibitem[Daraboina et~al.(2015)Daraboina, Pachitsas, and Solms]{Daraboina2015}
N.~Daraboina, S.~Pachitsas, and N.~V. Solms.
\newblock Natural gas hydrate formation and inhibition in gas/crude oil/aqueous
  systems.
\newblock \emph{Fuel}, 148:\penalty0 186--190, 2015.
\newblock ISSN 00162361.
\newblock \doi{10.1016/j.fuel.2015.01.103}.
\newblock URL \url{http://dx.doi.org/10.1016/j.fuel.2015.01.103}.

\bibitem[Davy(1811)]{Davy1811}
H.~Davy.
\newblock On some of the combinations of oxymuriatic gas andoxygene, and on the
  chemical relations of these principles, to inflammable bodies.
\newblock \emph{Philos. Trans. R. Soc}, 37:\penalty0 1--35, 1811.

\bibitem[Eslamimanesh et~al.(2012)Eslamimanesh, Mohammadi, Richon, Naidoo, and
  Ramjugernath]{Eslamimanesh2012a}
A.~Eslamimanesh, A.~H. Mohammadi, D.~Richon, P.~Naidoo, and D.~Ramjugernath.
\newblock Application of gas hydrate formation in separation processes: A
  review of experimental studies.
\newblock \emph{The Journal of Chemical Thermodynamics}, 46:\penalty0 62--71,
  2012.
\newblock ISSN 0021-9614.
\newblock \doi{https://doi.org/10.1016/j.jct.2011.10.006}.
\newblock URL
  \url{http://www.sciencedirect.com/science/article/pii/S0021961411003570}.

\bibitem[Fan et~al.(2009)Fan, Li, Wang, Lang, and Wang]{Fan2009a}
S.~Fan, S.~Li, J.~Wang, X.~Lang, and Y.~Wang.
\newblock Efficient capture of co2 from simulated flue gas by formation of tbab
  or tbaf semiclathrate hydrates.
\newblock \emph{Energy and Fuels}, 23:\penalty0 4202--4208, 2009.
\newblock ISSN 08870624.
\newblock \doi{10.1021/ef9003329}.

\bibitem[Gallo et~al.(2016)Gallo, Amann-Winkel, Angell, Anisimov, Caupin,
  Chakravarty, Lascaris, Loerting, Panagiotopoulos, Russo, Sellberg, Stanley,
  Tanaka, Vega, Xu, and Pettersson]{Gallo2016}
P.~Gallo, K.~Amann-Winkel, C.~A. Angell, M.~A. Anisimov, F.~Caupin,
  C.~Chakravarty, E.~Lascaris, T.~Loerting, A.~Z. Panagiotopoulos, J.~Russo,
  J.~A. Sellberg, H.~E. Stanley, H.~Tanaka, C.~Vega, L.~Xu, and L.~G.~M.
  Pettersson.
\newblock Water: A tale of two liquids.
\newblock \emph{Chemical Reviews}, 116:\penalty0 7463--7500, 7 2016.
\newblock \doi{10.1021/acs.chemrev.5b00750}.

\bibitem[Gudmundsson et~al.(1994)Gudmundsson, Parlaktuna, and
  Khokhar]{Gudmundsson1994}
J.~Gudmundsson, M.~Parlaktuna, and A.~Khokhar.
\newblock Storage of natural gas as frozen hydrate.
\newblock \emph{SPE Production \& Facilities}, 9:\penalty0 69--73, 1994.

\bibitem[Hammerschmidt(1934)]{Hammerschmidt1934}
E.~Hammerschmidt.
\newblock Formation of gas hydrates in natural gas transmission lines.
\newblock \emph{Ind. Eng. Chem.}, 26:\penalty0 851--855, 1934.

\bibitem[Heidaryan et~al.(2010)Heidaryan, Salarabadi, Moghadasi, and
  Dourbash]{Heidaryan2010}
E.~Heidaryan, A.~Salarabadi, J.~Moghadasi, and A.~Dourbash.
\newblock A new high performance gas hydrate inhibitor.
\newblock \emph{Journal of Natural Gas Chemistry}, 19:\penalty0 323--326, 2010.
\newblock ISSN 10039953.
\newblock \doi{10.1016/S1003-9953(09)60060-8}.
\newblock URL \url{http://dx.doi.org/10.1016/S1003-9953(09)60060-8}.

\bibitem[Horne and Johnson(1966)]{Horne1966}
R.~Horne and D.~Johnson.
\newblock The viscosity of water under pressure.
\newblock \emph{The Journal of Physical Chemistry}, 70:\penalty0 2182--2190, 5
  1966.
\newblock \doi{10.1021/j100879a018}.
\newblock URL \url{https://pubs.acs.org/sharingguidelines}.

\bibitem[Kang and Lee(2000)]{Kang2000a}
S.~P. Kang and H.~Lee.
\newblock Recovery of co2 from flue gas using gas hydrate: Thermodynamic
  verification through phase equilibrium measurements.
\newblock \emph{Environmental Science and Technology}, 34:\penalty0 4397--4400,
  2000.
\newblock ISSN 0013936X.
\newblock \doi{10.1021/es001148l}.

\bibitem[Krieger and Dougherty(1959)]{Krieger1959}
I.~M. Krieger and T.~J. Dougherty.
\newblock A mechanism for non‐newtonian flow in suspensions of rigid spheres.
\newblock \emph{Transactions of the Society of Rheology}, 3:\penalty0 137--152,
  3 1959.
\newblock ISSN 0038-0032.
\newblock \doi{10.1122/1.548848}.
\newblock URL \url{https://doi.org/10.1122/1.548848}.
\newblock doi: 10.1122/1.548848.

\bibitem[Linga et~al.(2007)Linga, Adeyemo, and Englezos]{Linga2007}
P.~Linga, A.~Adeyemo, and P.~Englezos.
\newblock Medium-pressure clathrate hydrate/membrane hybrid process for
  postcombustion capture of carbon dioxide.
\newblock \emph{Environmental Science \& Technology}, 42:\penalty0 315--320, 11
  2007.
\newblock \doi{10.1021/es071824k}.

\bibitem[Loekman et~al.(2019)Loekman, Claßen, Seidl, Luzi, Gatternig, Rauh,
  and Delgado]{Loekman2019}
S.~Loekman, T.~Claßen, P.~Seidl, G.~Luzi, B.~Gatternig, C.~Rauh, and
  A.~Delgado.
\newblock Potential application of innovative gas-hydrate technology in fruit
  juices concentration process.
\newblock \emph{2019 World Congress on Advances in Nano, Bio, Robotics, and
  Energy (ANBRE19)}, 2019.

\bibitem[Majid et~al.(2017)Majid, Wu, and Koh]{Majid2017}
A.~A. Majid, D.~T. Wu, and C.~A. Koh.
\newblock New in situ measurements of the viscosity of gas clathrate hydrate
  slurries formed from model water-in-oil emulsions.
\newblock \emph{Langmuir}, 33:\penalty0 11436--11445, 10 2017.
\newblock ISSN 15205827.
\newblock \doi{10.1021/acs.langmuir.7b02642}.

\bibitem[Majid et~al.(2018)Majid, Wu, and Koh]{Majid2018}
A.~A.~A. Majid, D.~T. Wu, and C.~A. Koh.
\newblock A perspective on rheological studies of gas hydrate slurry
  properties.
\newblock \emph{Engineering}, 4:\penalty0 321--329, 2018.
\newblock \doi{https://doi.org/10.1016/j.eng.2018.05.017}.
\newblock URL
  \url{http://www.sciencedirect.com/science/article/pii/S2095809917308202}.

\bibitem[McElligott et~al.(2019)McElligott, Uddin, Meunier, and
  Servio]{McElligott2019}
A.~McElligott, H.~Uddin, J.~L. Meunier, and P.~Servio.
\newblock Effects of hydrophobic and hydrophilic graphene nanoflakes on methane
  hydrate kinetics.
\newblock \emph{Energy and Fuels}, 33:\penalty0 11705--11711, 2019.
\newblock ISSN 15205029.
\newblock \doi{10.1021/acs.energyfuels.9b02927}.

\bibitem[McElligott et~al.(2021)McElligott, Meunier, and
  Servio]{McElligott2021a}
A.~McElligott, J.-L. Meunier, and P.~Servio.
\newblock Effects of hydrophobic and hydrophilic graphene nanoflakes on methane
  dissolution rates in water under vapor–liquid–hydrate equilibrium
  conditions.
\newblock \emph{Industrial and Engineering Chemistry Research}, 60:\penalty0
  2677--2685, 2 2021.
\newblock \doi{10.1021/acs.iecr.0c05808}.

\bibitem[Mills(1985)]{Mills1985}
P.~Mills.
\newblock Non-newtonian behaviour of flocculated suspensions.
\newblock \emph{Journal de Physique Lettres}, 46:\penalty0 301--309, 1985.
\newblock URL
  \url{https://hal.archives-ouvertes.fr/file/index/docid/232515/filename/ajp-jphyslet_1985_46_7_301_0.pdf}.

\bibitem[Mimachi et~al.(2015)Mimachi, Takahashi, Takeya, Gotoh, Yoneyama,
  Hyodo, Takeda, and Murayama]{Mimachi2015}
H.~Mimachi, M.~Takahashi, S.~Takeya, Y.~Gotoh, A.~Yoneyama, K.~Hyodo,
  T.~Takeda, and T.~Murayama.
\newblock Effect of long-term storage and thermal history on the gas content of
  natural gas hydrate pellets under ambient pressure.
\newblock \emph{Energy and Fuels}, 29:\penalty0 4827--4834, 7 2015.
\newblock \doi{10.1021/acs.energyfuels.5b00832}.

\bibitem[nam Park et~al.(2011)nam Park, Hong, Lee, Kang, Lee, Ha, and
  Lee]{Park2011}
K.~nam Park, S.~Y. Hong, J.~W. Lee, K.~C. Kang, Y.~C. Lee, M.-G. Ha, and J.~D.
  Lee.
\newblock A new apparatus for seawater desalination by gas hydrate process and
  removal characteristics of dissolved minerals (na+, mg2+, ca2+, k+, b3+).
\newblock \emph{Desalination}, 274:\penalty0 91--96, 2011.
\newblock ISSN 0011-9164.
\newblock \doi{https://doi.org/10.1016/j.desal.2011.01.084}.

\bibitem[Pandey and Sangwai(2020)]{Pandey2020}
G.~Pandey and J.~S. Sangwai.
\newblock High pressure rheological studies of methane hydrate slurries formed
  from water-hexane, water-heptane, and water-decane multiphase systems.
\newblock \emph{Journal of Natural Gas Science and Engineering}, 81:\penalty0
  103365, 2020.
\newblock ISSN 1875-5100.
\newblock \doi{https://doi.org/10.1016/j.jngse.2020.103365}.
\newblock URL
  \url{https://www.sciencedirect.com/science/article/pii/S1875510020302195}.

\bibitem[Pandey et~al.(2017)Pandey, Linga, and Sangwai]{Pandey2017}
G.~Pandey, P.~Linga, and J.~S. Sangwai.
\newblock High pressure rheology of gas hydrate formed from multiphase systems
  using modified couette rheometer.
\newblock \emph{Review of Scientific Instruments}, 88:\penalty0 25102, 2 2017.
\newblock ISSN 0034-6748.
\newblock \doi{10.1063/1.4974750}.
\newblock URL \url{https://doi.org/10.1063/1.4974750}.
\newblock doi: 10.1063/1.4974750.

\bibitem[Pasieka et~al.(2015)Pasieka, Jorge, Coulombe, and Servio]{Pasieka2015}
J.~Pasieka, L.~Jorge, S.~Coulombe, and P.~Servio.
\newblock Effects of as-produced and amine-functionalized multi-wall carbon
  nanotubes on carbon dioxide hydrate formation.
\newblock \emph{Energy and Fuels}, 29:\penalty0 5259--5266, 8 2015.
\newblock ISSN 0887-0624.
\newblock \doi{10.1021/acs.energyfuels.5b01036}.

\bibitem[Posteraro et~al.(2015{\natexlab{a}})Posteraro, Ivall, Maric, and
  Servio]{Posteraro2015_part2}
D.~Posteraro, J.~Ivall, M.~Maric, and P.~Servio.
\newblock New insights into the effect of polyvinylpyrrolidone (pvp)
  concentration on methane hydrate growth. 2. liquid phase methane mole
  fraction.
\newblock \emph{Chemical Engineering Science}, 126:\penalty0 91--98,
  2015{\natexlab{a}}.
\newblock ISSN 00092509.
\newblock \doi{10.1016/j.ces.2014.12.008}.
\newblock URL
  \url{https://www.sciencedirect.com/science/article/pii/S0009250914007222}.

\bibitem[Posteraro et~al.(2015{\natexlab{b}})Posteraro, Verrett, Maric, and
  Servio]{Posteraro2015_part1}
D.~Posteraro, J.~Verrett, M.~Maric, and P.~Servio.
\newblock New insights into the effect of polyvinylpyrrolidone (pvp)
  concentration on methane hydrate growth. 1. growth rate.
\newblock \emph{Chemical Engineering Science}, 126:\penalty0 99--105,
  2015{\natexlab{b}}.
\newblock ISSN 0009-2509.
\newblock \doi{https://doi.org/10.1016/j.ces.2014.12.009}.
\newblock URL
  \url{http://www.sciencedirect.com/science/article/pii/S0009250914007234}.

\bibitem[Rajput et~al.(2018)Rajput, Colantuoni, Bayahya, Dhane, Servio, and
  Maric]{Rajput2018}
F.~Rajput, A.~Colantuoni, S.~Bayahya, R.~Dhane, P.~Servio, and M.~Maric.
\newblock Poly(styrene/pentafluorostyrene)-block-poly(vinyl
  alcohol/vinylpyrrolidone) amphiphilic block copolymers for kinetic gas
  hydrate inhibitors: Synthesis, micellization behavior, and methane hydrate
  kinetic inhibition.
\newblock \emph{Journal of Polymer Science Part A: Polymer Chemistry},
  56:\penalty0 2445--2457, 11 2018.
\newblock ISSN 0887-624X.
\newblock \doi{https://doi.org/10.1002/pola.29219}.
\newblock URL \url{https://doi.org/10.1002/pola.29219}.
\newblock https://doi.org/10.1002/pola.29219.

\bibitem[Rajput et~al.(2021)Rajput, Maric, and Servio]{Rajput2021}
F.~Rajput, M.~Maric, and P.~Servio.
\newblock Amphiphilic block copolymers with vinyl caprolactam as kinetic gas
  hydrate inhibitors.
\newblock \emph{Energies}, 14, 2021.
\newblock \doi{10.3390/en14020341}.

\bibitem[Servio and Englezos(2001)]{Servio2001}
P.~Servio and P.~Englezos.
\newblock Effect of temperature and pressure on the solubility of carbon
  dioxide in water in the presence of gas hydrate.
\newblock \emph{Fluid Phase Equilibria}, 190:\penalty0 127--134, 11 2001.
\newblock ISSN 03783812.
\newblock \doi{10.1016/S0378-3812(01)00598-2}.
\newblock URL
  \url{https://www.sciencedirect.com/science/article/pii/S0378381201005982
  https://linkinghub.elsevier.com/retrieve/pii/S0378381201005982}.

\bibitem[Servio and Englezos(2002)]{Servio2002}
P.~Servio and P.~Englezos.
\newblock Measurement of dissolved methane in water in equilibrium with its
  hydrate.
\newblock \emph{Journal of Chemical \& Engineering Data}, 47:\penalty0 87--90,
  1 2002.
\newblock ISSN 0021-9568.
\newblock \doi{10.1021/je0102255}.
\newblock URL \url{https://doi.org/10.1021/je0102255}.
\newblock doi: 10.1021/je0102255.

\bibitem[Sloan and Koh(2008)]{Sloan2008}
E.~Sloan and C.~Koh.
\newblock \emph{Clathrate hydrates of natural gases}.
\newblock Taylor and Francis, 3rd edition, 2008.

\bibitem[Sun et~al.(2020)Sun, Fu, Wang, Xu, Chen, Wang, and Zhang]{Sun2020}
B.~Sun, W.~Fu, Z.~Wang, J.~Xu, L.~Chen, J.~Wang, and J.~Zhang.
\newblock Characterizing the rheology of methane hydrate slurry in a horizontal
  water-continuous system.
\newblock \emph{SPE Journal}, 25:\penalty0 1026--1041, 6 2020.
\newblock ISSN 1086-055X.
\newblock \doi{10.2118/195586-PA}.
\newblock URL \url{https://doi.org/10.2118/195586-PA}.

\bibitem[Webb et~al.(2012)Webb, Rensing, Koh, Sloan, Sum, and
  Liberatore]{Webb2012}
E.~B. Webb, P.~J. Rensing, C.~A. Koh, E.~D. Sloan, A.~K. Sum, and M.~W.
  Liberatore.
\newblock High-pressure rheology of hydrate slurries formed from water-in-oil
  emulsions.
\newblock \emph{Energy and Fuels}, 26:\penalty0 3504--3509, 6 2012.
\newblock ISSN 0887-0624.
\newblock \doi{10.1021/ef300163y}.
\newblock URL \url{https://doi.org/10.1021/ef300163y}.
\newblock doi: 10.1021/ef300163y.

\bibitem[Webb et~al.(2013)Webb, Koh, and Liberatore]{Webb2013}
E.~B. Webb, C.~A. Koh, and M.~W. Liberatore.
\newblock Rheological properties of methane hydrate slurries formed from aot +
  water + oil microemulsions.
\newblock \emph{Langmuir}, 29:\penalty0 10997--11004, 9 2013.
\newblock ISSN 0743-7463.
\newblock \doi{10.1021/la4022432}.
\newblock URL \url{https://doi.org/10.1021/la4022432}.
\newblock doi: 10.1021/la4022432.

\bibitem[Webb et~al.(2014)Webb, Koh, and Liberatore]{Webb2014}
E.~B. Webb, C.~A. Koh, and M.~W. Liberatore.
\newblock High pressure rheology of hydrate slurries formed from
  water-in-mineral oil emulsions.
\newblock \emph{Industrial and Engineering Chemistry Research}, 53:\penalty0
  6998--7007, 4 2014.
\newblock \doi{10.1021/ie5008954}.

\bibitem[Wonnham(1967)]{Wonnham1967}
J.~Wonnham.
\newblock Effect of pressure on the viscosity of water.
\newblock \emph{Nature}, 215:\penalty0 1053--1054, 1967.
\newblock \doi{10.1038/2151053a0}.
\newblock URL \url{https://doi.org/10.1038/2151053a0}.

\bibitem[Zhukov et~al.(2017)Zhukov, Stolov, and Varfolomeev]{Zhukov2017a}
A.~Y. Zhukov, M.~A. Stolov, and M.~A. Varfolomeev.
\newblock Use of kinetic inhibitors of gas hydrate formation in oil and gas
  production processes: Current state and prospects of development.
\newblock \emph{Chemistry and Technology of Fuels and Oils}, 53:\penalty0
  377--381, 2017.
\newblock ISSN 15738310.
\newblock \doi{10.1007/s10553-017-0814-6}.

\end{thebibliography}

\section*{Author Contributions}
The manuscript was written through the contributions of all authors. All authors have approved the final version of the manuscript.

\section*{Declaration of Competing Interest}
The authors declare that they have no known competing financial interests or personal relationships that could have appeared to influence the work reported in this paper.

\section*{Acknowledgements}
The authors would like to acknowledge the financial support from the Natural Sciences and Engineering Research Council of Canada (NSERC) grant number 206269, and from the McGill Engineering Doctoral Award (MEDA).


\clearpage
\beginsupplement

\section*{Supplementary Materials}
Here, we provide all supplementary materials used in our analysis. 

\end{document}